\documentclass[prb,twocolumn,showpacs,amsmath,amssymb,superscriptaddress]{revtex4-2}
\usepackage[pdftex]{hyperref,graphicx}
\hypersetup{colorlinks = true, urlcolor = blue, linkcolor = blue, citecolor = blue}
\usepackage{xcolor}
\usepackage{mathtools}
\usepackage{bm} 
\usepackage{dcolumn} 
\usepackage{graphicx} 
\usepackage{amsmath} 
\usepackage{amssymb} 
\usepackage{amsfonts} 
\usepackage{placeins}
\usepackage[caption=false]{subfig}
\usepackage{subfig}

\newcommand{\mdos}[1]{$m_\mathrm{DOS}$}

\begin{document}
\title{Fragile versus stable two-dimensional fermionic quasiparticles}

\author{Seongjin Ahn}
\affiliation{Condensed Matter Theory Center and Joint Quantum Institute, Department of Physics, University of Maryland, College Park, Maryland 20742-4111, USA}
\author{Sankar Das Sarma}
\affiliation{Condensed Matter Theory Center and Joint Quantum Institute, Department of Physics, University of Maryland, College Park, Maryland 20742-4111, USA}

\date{\today}

\begin{abstract}
We provide a comprehensive theoretical investigation of the Fermi liquid quasiparticle description in two-dimensional electron gas interacting via the long-range Coulomb interaction by calculating the electron self-energy within the leading-order approximation, which is exact in the high-density limit. We find that the quasiparticle energy is larger than the imaginary part of the self-energy up to very high energies, implying that the basic Landau quasiparticle picture is robust up to far above the Fermi energy. We find, however, that the quasiparticle picture becomes fragile in a small discrete region around a critical wave vector where the quasiparticle spectral function strongly deviates from the expected quasiparticle Lorentzian line shape with a vanishing renormalization factor. We show that such a non-Fermi liquid behavior arises due to the coupling of quasiparticles with the collective plasmon mode. This situation is somewhat intermediate between the one-dimensional interacting electron gas (i.e., Luttinger liquid), where the Landau Fermi liquid theory completely breaks down since only bosonic collective excitations exist, and three-dimensional electron gas, where quasiparticles are well-defined and more stable against interactions than in one and two dimensions. We use a number of complementary definitions for a quasiparticle to examine the interacting spectral function, contrasting two-dimensional and three-dimensional situations critically.
\end{abstract}

\maketitle

\section{Introduction} \label{sec:Introduction}
Landau Fermi liquid theory is one of the greatest triumphs of solid-state physics since it drastically simplifies interacting many fermion problems into the much simpler single-particle Fermi gas problems \cite{Landau1957, Landau1959, Baym2008Landau, Pines2018Theory}. Landau's idea is that the low-energy excitations of the interacting system have a one-to-one correspondence to their noninteracting counterparts, and thus behave as quasiparticles, which are similar to noninteracting particles in many ways but with renormalized effective mass and finite lifetime \cite{quinn1958electron, Rice1965, abrikosov2012methods, Jalabert1989, fetter2012quantum}. As long as the lifetime is long, quasiparticles are stable, long-lived, and well-defined. This enables us to continue to use, even in the presence of interactions, the concept of the energy band theory to describe the properties of metals and semiconductors, even for the excited states as long as the quasiparticle picture and the Landau Fermi liquid theory apply. In fact, much of the electronic solid-state physics is based on the validity of the Fermi liquid theory where the properties of metals and semiconductors are described in terms of quasielectrons and quasiholes, which, for all practical purposes, behave as noninteracting electrons and holes of a Fermi gas. 

There is a subtle point about the Fermi liquid theory which is not often emphasized.  The basic idea underlying the Fermi liquid theory is that the Fermi surface in the form of a discontinuity in the momentum distribution function survives in the interacting ground state, thus creating a one-to-one correspondence between the interacting Fermi liquid ground state and the non-interacting Fermi gas ground state, i.e., the Fermi liquid ground state is a fixed point of the interacting system.  This, however, is a statement on the ground state and does not say anything about quasiparticles except to imply that there must be zero energy gapless excitations around the Fermi surface even in the interacting Fermi liquid.  Thus, finite energy long-lived quasiparticles, carrying a renormalized effective mass and behaving like noninteracting particle-hole excitations, is a stronger statement than just the mere existence of a Fermi liquid fixed point as the interacting ground state. Our interest in the current paper is the nature of the quasiparticles at finite excitation energies, not just the existence of an interacting Fermi liquid fixed point, which has already been rigorously established \cite{feldman2004twoPart1, feldman2004twoPart2, feldman2004twoPart3}.

A question about the regime of validity of the Landau Fermi liquid theory has always been of central importance in condensed matter theory. The obvious condition for the Landau Fermi liquid to work is that the interacting ground state should be adiabatically connected to the noninteracting ground state so that the one-to-one correspondence between the interacting and noninteracting eigenstates holds. An example of the breakdown of the Landau Fermi liquid theory is an interacting one-dimensional electron gas (i.e., Luttinger liquid), where all the elementary excitations have bosonic collective nature because of the dimensionality of the system and there are no low energy particle-hole type single-particle excitations, thus destroying the Fermi surface as a stable fixed point of the interacting system \cite{Tomonaga1950, Luttinger1963}. Because of the central importance of the Fermi liquid theory in condensed matter physics, its possible breakdown is of great fundamental interest, and a terminology called `non-Fermi-liquids' has developed in describing interacting fermionic systems where the Fermi liquid paradigm may break down.  Some examples of such non-Fermi liquids are one-dimensional fermions, fractional quantum Hall liquids, spin liquids, Mott insulators, and quantum critical systems.  Our goal in the current work is focused not on the question of a non-Fermi-liquid, but on the question of the regime of validity (in energy and momentum) of the quasiparticle picture for a two-dimensional electron system interacting via the long-range Coulomb interaction.  The ground state of this system is certainly a Fermi liquid for weak interactions, but we want to know the extent to which the quasiparticle picture applies, i.e., how far from the Fermi surface can we still describe the interacting system in terms of quasiparticles?

The key condition for the Fermi liquid theory to apply is that quasiparticles should have long enough lifetime not to decay during the adiabatic switching process [see Eq.~(\ref{eq:quasiparticle_condition}) in Sec.~\ref{sec:quasiparticle}]. It is well known that quasiparticles are infinitely long lived at the Fermi surface (i.e. gapless zero energy excitations) with the inverse of its lifetime scaling as $\sim E^2_\mathrm{QP}\ln{E_\mathrm{QP}}$ \cite{Zheng1996} and $\sim E^2_\mathrm{QP}$  for two-dimensional (2D) and three-dimensional electron gas (3DEG), respectively, where $E_\mathrm{QP}$ is the quasiparticle energy measured from the Fermi surface. In fact, this perturbative result for the behavior of quasiparticle decay at the Fermi surface can be formalized into a renormalization group flow argument to establish the perturbative existence of a Fermi liquid fixed point in 2D and 3D systems, with similar arguments also establishing the nonexistence of the Fermi liquid fixed point in 1D interacting systems \cite{Polchinski1992Effective, Shankar1994Renormalization}. Thus, quasiparticles are usually considered to be well-defined only in the immediate vicinity of the Fermi surface, and become ill-defined at high energies away from the Fermi surface. In fact, it is universally assumed that the validity of the quasiparticle picture necessarily requires $E_\mathrm{QP}\ll E_F$, i.e., quasiparticles exist only at low excitation energies and, consequently, at momenta not too far from the Fermi momenta. A natural question that arises is then: What is the energy range of validity of the Fermi liquid theory? How far away from the Fermi surface can we go and still use long-lived and well-defined quasiparticles as the dominant excitations of the interacting system? This question, to the best of our knowledge, has not been addressed with the level of quantitative seriousness it deserves despite the importance of the concept of quasiparticle in condensed matter theory.
In fact, quasiparticles far away from the Fermi surface are often assumed to be well-defined purely on empirical basis without any theoretical justification. There is also a widespread counter-opinion that the quasiparticle picture absolutely necessitates the low energy condition, $E_\mathrm{QP}\ll E_F$, and quasiparticles do not exist at high energies.

In this work, we directly address this issue by explicitly calculating the self-energy and the spectral function of interacting electrons up to high energies of several orders of the Fermi energy. We use the leading-order dynamical approximation in the many-body perturbation theory for the Coulomb interacting electron systems \cite{GellMann1957}, which has been widely successful in describing quasiparticle properties in various materials. The theory is well-controlled and exact in the high-density limit, where the Coulomb interaction is perturbatively weak. Even though the leading-order dynamical screening theory is exact in the weakly interacting limit, the theory has been shown to be reasonably valid even in the metallic range of density \cite{Rice1965}.

The main goal of this paper is to provide a comprehensive study on the Fermi liquid theory for long-range Coulomb interactions with an in-depth discussion of its validity at high energies, extending substantially the previous Fermi liquid studies which are limited only to low energy regime near the Fermi surface \cite{Landau1957, Landau1959, Baym2008Landau, Pines2018Theory}. Our results show that the quasiparticle picture is robust up to high energies, explaining why the single-particle framework has been so successful in metals and semiconductors.
However, we find some clear signals of fragile 2D quasiparticles in a small region around a certain critical wave vector because of the coupling of quasiparticles with the collective plasmon excitations, which are often neglected in the discussion of the validity of the Fermi liquid theory in the existing literature. Plasmons are specific features of long-range Coulomb interactions, and they are gapped (ungapped) in 3(2)D, making their coupling to the quasiparticles of particular importance in 2D electron systems, making 2D quasiparticles more fragile than their 3D counterparts.

The paper is organized as follows. In Sec.~\ref{sec:theory}, we briefly review the Landau quasiparticle theory and introduce the formalism for the self-energy calculations within the random phase approximation (RPA) for the interacting self-energy. 
In Sec.~\ref{sec:Validity_of_quasiparticle_picture}, we investigate the validity of the quasiparticle picture for interacting 2DEG up to high energies of the order of several Fermi energies by explicitly comparing the quasiparticle energy with the imaginary part of the self-energy defining the quasiparticle decay rate. 
In Sec.~\ref{sec:spectral_function}, we present the calculated self-energy and spectral functions, showing that the spectral quasiparticle peak exhibits a non-Fermi-liquid-like behavior around a certain critical wave vector $k_c$ where the quasiparticle picture becomes suspect and fragile.
In Sec.~\ref{sec:renormalization_factor}, we provide further analysis of the anomalous quasiparticle behavior around the critical wave vector $k_c$ by evaluating several many-body quantities such as the renormalization factor $Z_k$, the effective mass $m_k$. We demonstrate that the fragility of quasiparticles is due to the coupling between quasiparticles and plasmon collective excitations.
In Sec.~\ref{sec:comparison_to_3DEG}, we compare the obtained 2DEG results with those for 3DEG. We show that the non-Fermi liquid features we find for 2DEG also appear similarly in 3DEG, but are much less prominent.
Section~\ref{sec:discussion_and_conclusion} contains the discussion and a summary.
We use $\hbar=1$ in the figures so that momentum/wave vector and energy/frequency are the same in our notations.  Also, we characterize the system by the Coulomb interaction strength $r_s$ throughout with $r_s$ small (large) being high-density weakly (low-density strongly) interacting system, and our theory is perturbatively exact in the high-density limit. 
Additional results for more $r_s$ values are provided in Appendix.


\section{Theory} \label{sec:theory}
\subsection{Quasiparticle }  \label{sec:quasiparticle}
The Landau quasiparticle theory was originally conceived based on phenomenological approach. Later on, a further development was made using the diagrammatic perturbative many-body formalism, reproducing the main results obtained on the phenomenological basis \cite{abrikosov2012methods}. Here, we briefly review the key concepts of the Landau quasiparticle theory in the perturbative many-body scheme. Landau's original work focused on short-range interactions as appropriate for 3D neutral fermions, e.g., normal He$^3$. Our focus is on 2D electrons interacting via the long-range Coulomb interaction in a positive jellium charge background maintaining charge neutrality.

Consider a quantum system of noninteracting fermions in the ground state characterized by a set of occupation numbers $n_{ k}=\Theta(k_\mathrm{F} - k)$. Suppose we add a fermion above the Fermi surface. The key idea of the Landau quasiparticle theory is that even in the presence of interactions the added fermion continues to behave as a single-particle excitation dressed by interactions, i.e., a quasiparticle, which is not an exact eigenstate of the interacting system, but is an almost eigenstate with a very long lifetime. It is clear that close to the Fermi surface, the quasiparticle decay is severely restricted because of the Pauli principle, and in fact, right on the Fermi surface the quasiparticles are infinitely stable, preserving the Fermi surface in the interacting system. The question that still remains open, however, is how large the momentum k can be for such a quasiparticle to be well-defined. This is the issue we address. Landau described the quasiparticle picture through the adiabatic process: if one turns on interactions infinitely slowly through $V(t)=Ve^{-\frac{\eta}{\hbar}|t|}$ from $t=-\infty$, where $\eta$ is infinitesimally small, then the noninteracting states would evolve smoothly into the real interacting states, establishing one to one correspondence between the eigenstates of the interacting system and those of the original noninteracting counterpart. During the adiabatic evolution, the ground state is stable unless there is a level crossing between states, which leads to a phase transition of the system. The quasiparticle (i.e., the added fermion dressed by interactions), however, undergoes a decay due to its interaction with other particles, and thus has a finite lifetime. The adiabatic process is irreversible if the interaction is not fully turned on before the quasiparticle decays because of its finite lifetime. Thus, for the quasiparticle picture to make sense, the quasiparticle lifetime should be much longer than the time taken for the interaction to be fully turned on, i.e., $\frac{\hbar}{\tau}\ll\eta$
where $\tau$ denotes the quasiparticle lifetime.
It is also required that interaction grows slowly enough so that the excited quasiparticle state is not mixed with other states during the adiabatic process, i.e., $\eta \ll E_\mathrm{QP}$, where $E_\mathrm{QP}$ is the quasiparticle energy.
These two requirements together give the condition for the quasiparticle picture to make sense:
\begin{equation}
    \frac{\hbar}{\tau} \ll E_\mathrm{QP},
    \label{eq:quasiparticle_condition}
\end{equation}
which is the standard Landau criterion for the existence of well-defined quasiparticles \cite{abrikosov2012methods}. There are, however, alternate criteria for defining quasiparticles which are extensively used, and the equivalence of various criteria for quasiparticles are not necessarily quantitatively equivalent although they all provide the same result on the Fermi surface where the quasiparticles are infinitely stable as gapless excitations.

Another criterion often used to determine the validity of the quasiparticle picture is to look at the shape of the spectral function and its broadening induced by interactions. The spectral function is written as
\begin{equation}
    A(k,\omega)= -\frac{1}{\pi}\mathrm{Im}G(k,\omega)
    \label{eq:spectral_function}
\end{equation}
and can be directly measured by experiments such as angle-resolved photoemission spectroscopy (ARPES).
Here $G(k,\omega)$ is the interacting Green's function given by 
\begin{equation}
    G^{-1}(k,\omega)= G_0^{-1}(k,\omega) +  \Sigma(k,\omega) 
    \label{eq:interacting_green}
\end{equation}
where $G_0^{-1}(k,\omega)$ is the bare noninteracting Green's function and
$\Sigma(k,\omega)$ denotes the self-energy, encoding all the effects arising from interactions.
For noninteracting systems, the Green's functions is given by 
\begin{equation}
    G_0(k,\omega)=\frac{1}{\hbar\omega-\xi_k+i\eta}
    \label{eq:noninteracting_green}
\end{equation}
with $\eta=|\eta| \mathrm{sgn}(\varepsilon_k)$ being infinitesimally small, where $\xi_k=\varepsilon_k - E_\mathrm{F}$ is the usual parabolic energy dispersion measured from the Fermi energy $E_\mathrm{F}$. Thus, the noninteracting spectral function is given by a $\delta$-function $A_0(k,\omega)=\delta(\hbar\omega-\xi_k)$. The $\delta$-function shape for the noninteracting spectral function means that when one adds a fermion, it occupies an exact eigenstate of the system with a precise relationship between energy and momentum since the noninteracting stationary states are also momentum eigenstates for the noninteracting free particle wavefunctions. Rather amazingly, the interacting ground state with $k=k_\mathrm{F}$ also has a $\delta$-function piece, reflecting the preservation of the Fermi surface in the presence of interactions. On the other hand, a quasiparticle state for $k>k_\mathrm{F}$ in an interacting system is not an exact eigenstate, thus decaying over a finite lifetime. This results in uncertainty in the quasiparticle energy due to the time-energy uncertainty principle, which simply means that a quasiparticle state does not have a sharp well-defined energy. Thus, quasiparticles typically appear (except for $k=k_F$) as a Lorentzian peak in the spectral function with a finite width determined through $\eta \sim \hbar/\tau$ in Eq.~(\ref{eq:noninteracting_green}) and its center being around the renormalized quasiparticle energy $E_\mathrm{QP}$. It is worth emphasizing that in this context Eq.~(\ref{eq:quasiparticle_condition}) can be interpreted as requiring the uncertainty in energy ($\hbar/\tau$) to be larger than the quasiparticle energy itself ($E_\mathrm{QP}$). 
Basically, Eq.~(\ref{eq:quasiparticle_condition}) asserts that a quasiparticle is ill-defined when the energy-broadening of the state is larger than the energy of the state itself, which is precisely what one infers also from the interacting spectral function.
Thus, the quasiparticle spectral peak becomes more broadened (and suppressed) as the corresponding quasiparticle becomes more fragile, and eventually the spectral peak vanishes when the Landau quasiparticle picture breaks down.
Another consequence of quasiparticles becoming fragile is that the spectral quasiparticle peak becomes increasingly non-Lorentzian, strongly deviating from the single-particle spectral function, which is Lorentzian as discussed above. We emphasize that a strong deviation from the Lorentzian shape for the interacting spectral function implies that the quasiparticle picture cannot solely describe the whole system even in the case where the spectral function exhibits a sharp but non-Lorentzian peak structure. 
In the literature, one sometimes loosely refers to the sharp Lorentzian (or $\delta$-function)-like spectral peak with small level broadening as the ``coherent" peak representing the well-defined quasiparticle, and the rest of the broadened featureless spectral function as ``incoherent" representing the non-quasiparticle part of the interacting excitations. As long as the interacting system manifests a well-defined and sharp Lorentzian spectral peak (with the broadening less than the peak energy), one can talk about quasiparticles. On the Fermi surface, this coherent quasiparticle peak becomes perfectly coherent with infinite lifetime since the broadening $\eta$ vanishes leading to a $\delta$-function describing the quasiparticle spectral function with zero excitation energy. For any finite $E_\mathrm{QP}$, there must always be a finite broadening $\eta$ (and hence a finite lifetime $\tau \sim \hbar/\eta$), and quasiparticles are stable and well-defined as long as $E_\mathrm{QP}\gg \eta$.


\subsection{Leading-order theory}
Having reviewed the concept of the Fermi liquid theory, in this section we introduce how we obtain the interacting spectral function given by Eq.~(\ref{eq:spectral_function}).
In this work, we use the leading-order dynamical approximation to evaluate the self-energy \cite{abrikosov2012methods}. 
The leading-order theory for Coulomb interaction involves an expansion of the interacting self-energy in terms of the dynamically screened Coulomb interaction where an infinite series of ring diagrams is inserted in the Coulomb propagator allowing vacuum polarization by electron-hole bubbles. This ring diagram series includes the leading Coulomb divergence in each order, and is convergent upon resummation. The theory is known to be exact in the high-density limit, and is the only controlled analytical theory available for the electron liquid interacting via the long-range Coulomb interaction. The pure Hartree tadpole diagrams vanish by virtue of charge neutrality, and one needs to resum the infinite ring diagrams. This is often called RPA in the literature, a terminology we use extensively in this paper from now on-- all it means in our context is that the self-energy keeps the exchange energy and the infinite series of ring diagrams in the dynamically screened Coulomb interaction. The theory is exact in the high-density limit, and is thus a well-defined many-body approximation. 
This theory is sometimes dubbed as ``$G_0W_0$ approximation" in the Hedin's $GW$ framework\cite{Hedin1965New} with $G_0$ indicating the noninteracting Green's function and $W_0$ the dynamically screened Coulomb interaction within RPA. Within the $GW$ scheme, one can go beyond the leading order approximation by using the full interacting Green's function $G$, solving the fully self-consistent $GW$ equations. The fully self-consistent $GW$ approximation, however, is rarely used since it is numerically demanding and generally gives results worse than the leading order approximation because it badly mixes perturbative orders in an uncontrolled way with the vertex correction neglected. Thus in the following we use the terminology ``RPA" instead of ``$GW$" to emphasize the leading order nature of our calculations, which are exact in the high density limit.
We emphasize that in spite of its widespread use in band structure calculations, the so-called ``$GW$ approximation" is not a consistent perturbative many-body approximation because of mixing orders, and what we use, the so-called ``$G_0W_0$" approximation or the RPA theory is the appropriate leading order approximation.
Strictly speaking, the RPA approximation is valid only in the weakly interacting limit where $r_s\ll1$ and $r_s$ is the interaction strength (i.e., the dimensionless Wigner-Seitz radius providing the average inter-electron separation in units of the effective Bohr radius-- a high or low density system with $r_s<1$ or $r_s>1$ is weakly or strongly interacting respectively) defined through $n^{-1}=\pi(r_s a_\mathrm{B})^2$ where $a_B$ is the Bohr radius and $n$ is the electron density \cite{mahan2000many}. 
In 3D, $r_s$ is defined by $n^{-1}= 4\pi(r_s a_\mathrm{B})^3/3$ where $n$ is now the 3D electron density. We note that $r_s$ is simply the inter-particle average separation measured in the atomic Bohr radius units so that large (small) $r_s$ is respectively the small (large) density limit. Thus, $r_s$ is obtained by knowing the carrier density of the system.
However, it has been empirically shown that even for large $r_s$ up to $6$ the leading-order dynamical approximation is highly successful in predicting quasiparticle properties of various systems including 2DEG and 3DEG. Thus, in this work we present results obtained for both small and large $r_s$, and show that the important qualitative features remain the same regardless of the value of $r_s$. We emphasize, however, that our results are strictly valid only for small $r_s$($<1$) where our leading-order dynamically screened RPA theory is perturbatively exact, and our large $r_s$ results in the strongly interacting regime are given only for the sake of completeness. 
We note that, as is well-known, the interacting electron liquid Hamiltonian, containing the noninteracting kinetic energy and the Coulomb interaction, can be easily cast into the dimensionless form containing only $r_s$, with the kinetic energy being $O( 1/ r_s^2)$ and the interaction energy being $O( 1/r_s)$, showing that the system is weakly interacting for $r_s<1$ \cite{quinn1958electron, Rice1965, abrikosov2012methods, Jalabert1989}.

Within the leading-order dynamical approximation, the electron self-energy is given by
\begin{align}
\Sigma({\bm k},i\omega_n)\!=&-\!\int\!\frac{d^2 q}{(2\pi)^2}\frac{1}{\beta}\sum_{i\Omega_n}W(\bm q,i\Omega_n) \nonumber \\ 
&\times G_0(\bm k + \bm q, i\omega_n+i\Omega_n),
\label{eq:GW_self_energy}
\end{align}
where $\omega_n$ and $\Omega_n$ are Matsubara frequencies, $G_0=\left(i\hbar\omega_n+i\Omega_n-\xi_{\bm k + \bm q} \right)^{-1}$ is the noninteracting Green's function, $\beta=(k_\mathrm{B}T)^{-1}$, $T$ is the temperature, $k_\mathrm{B}$ is the Boltzmann constant and 
\begin{equation}
    W(\bm q,i\Omega_n)=v_c(\bm q)/\varepsilon(q,i\Omega_n)    
    \label{eq:screened_Coulomb}
\end{equation}
denotes the RPA dynamically screened Coulomb interaction where $v_c(\bm q)=2\pi e^2/|\bm q|$ is the two-dimensional bare Coulomb interaction, and the denominator in Eq.~(\ref{eq:screened_Coulomb}) arises from the summing of the infinite series of electron-hole ring diagrams forming a geometric series.
Here the dielectric function
\begin{equation}
    \varepsilon(q, \omega)=1-v_c(\bm q)\Pi_0(\bm q,\omega)
    \label{eq:diel}
\end{equation}
is obtained within the RPA with $\Pi_0(\bm q,\omega)$ being the noninteracting polarization function given by \cite{Stern1967}
\begin{align}
    \Pi_0(\bm q,\omega)=&-\frac{m}{\pi} + \frac{m^2}{\pi q^2}
    \left[
    \sqrt{\left(    \omega+\frac{q^2}{2m}   \right)^2-\frac{2E^0_\mathrm{F} q^2}{2m}}\right.\nonumber \\
    &-
    \left.\sqrt{\left(    \omega-\frac{q^2}{2m}   \right)^2-\frac{2E^0_\mathrm{F} q^2}{2m}}\right],
    \label{eq:iso_polar}
\end{align}
where $E^0_\mathrm{F}$ is the bare Fermi energy.

The self-energy given by Eq.~(\ref{eq:GW_self_energy}) can be decomposed into the static exchange and correlation parts: $\Sigma=\Sigma^\mathrm{ex}+\Sigma^\mathrm{corr}$, where 
$\Sigma^\mathrm{ex}$ is the exchange self-energy given by 
\begin{align}
\Sigma^{\mathrm{ex}} ({\bm k})\!=&-\!\int\!\frac{d^2 q}{(2\pi)^2} \Theta(-\xi_{\bm k+\bm q}) v_c(\bm q)
\label{eq:self_energy_ex}
\end{align}
in the zero-temperature limit.
$\Sigma^\mathrm{corr}$ is the dynamical correlation part containing all contributions beyond exchange interaction written as
\begin{align}
    \Sigma^\mathrm{corr}({\bm k},i\omega_n)\!=&-\!\int\!\frac{d^2 q}{(2\pi)^2}\frac{1}{\beta}\sum_{i\Omega_n}
    \left[\frac{1}{\varepsilon(\bm q,i \Omega_n)}-1\right]
    \nonumber \\ 
&\times G_0(\bm k + \bm q, i\omega_n+i\Omega_n).    
\end{align}
It is useful to express the correlation self-energy as a sum of two terms: 
$\Sigma^\mathrm{corr}=\Sigma^\mathrm{line}+\Sigma^\mathrm{res}$.
The line part $\Sigma^\mathrm{line}$ is obtained by first performing an analytic continuation (i.e., $i\omega_n\rightarrow\omega+i\eta$) and then doing the Matsubara summation. In the zero temperature limit, we obtain 
\begin{align}
\Sigma^{\mathrm{line}} ({\bm k},\omega)\!=&-\!\int\!\frac{d^2 q}{(2\pi)^2}\int_{-\infty}^{\infty}\!\frac{d\Omega}{2\pi} 
\frac{v_c(\bm q)}{\xi_{\bm k+\bm q}-\hbar\omega-i\Omega} \nonumber \\ 
&\times\left[\frac{1}{\varepsilon(\bm q,i \Omega)}-1\right].
\label{eq:self_energy_line}
\end{align}
Note that the Matsubara summation is supposed to be done before performing an analytic continuation, and thus $\Sigma^{\mathrm{line}}$ is not the entire correlation self-energy. $\Sigma^{\mathrm{res}}$ gives the remaining part of the correlation self-energy, written as
\begin{align}
\Sigma^{\mathrm{res}} ({\bm k},\omega)\!=&\!\int\!\frac{d^2 q}{(2\pi)^2} \left [\Theta(\hbar\omega-\xi_{\bm k+\bm q}) - \Theta(-\xi_{\bm k+\bm q}) \right ] \nonumber \\ 
&\times v_c(\bm q)\left[\frac{1}{\varepsilon(\bm q,\xi_{\bm k+\bm q}-\hbar\omega)}-1\right].
\label{eq:self_energy_res}
\end{align}
It is easy to see that $\Sigma^\mathrm{line}$ is always real since $\varepsilon^*(\bm q,i\nu)=\varepsilon(\bm q,-i \nu)=\varepsilon(\bm q,i \nu)$. As $\Sigma^\mathrm{ex}$ ia real, which can be obviously seen from from Eq.~(\ref{eq:self_energy_ex}), the imaginary part of the self-energy is entirely determined by $\Sigma^\mathrm{res}$, thus
\begin{align}
    \mathrm{Im}\Sigma(\bm k, \omega)
    \!=&\!\int\!\frac{d^2 q}{(2\pi)^2} \left [\Theta(\hbar\omega-\xi_{\bm k+\bm q}) - \Theta(-\xi_{\bm k+\bm q}) \right ] \nonumber \\ 
&\times v_c(\bm q)\mathrm{Im}\left[\frac{1}{\varepsilon(\bm q,\xi_{\bm k+\bm q}-\hbar\omega)}\right].
\end{align}

\begin{figure}[!htb]
  \centering
  \includegraphics[width=\linewidth]{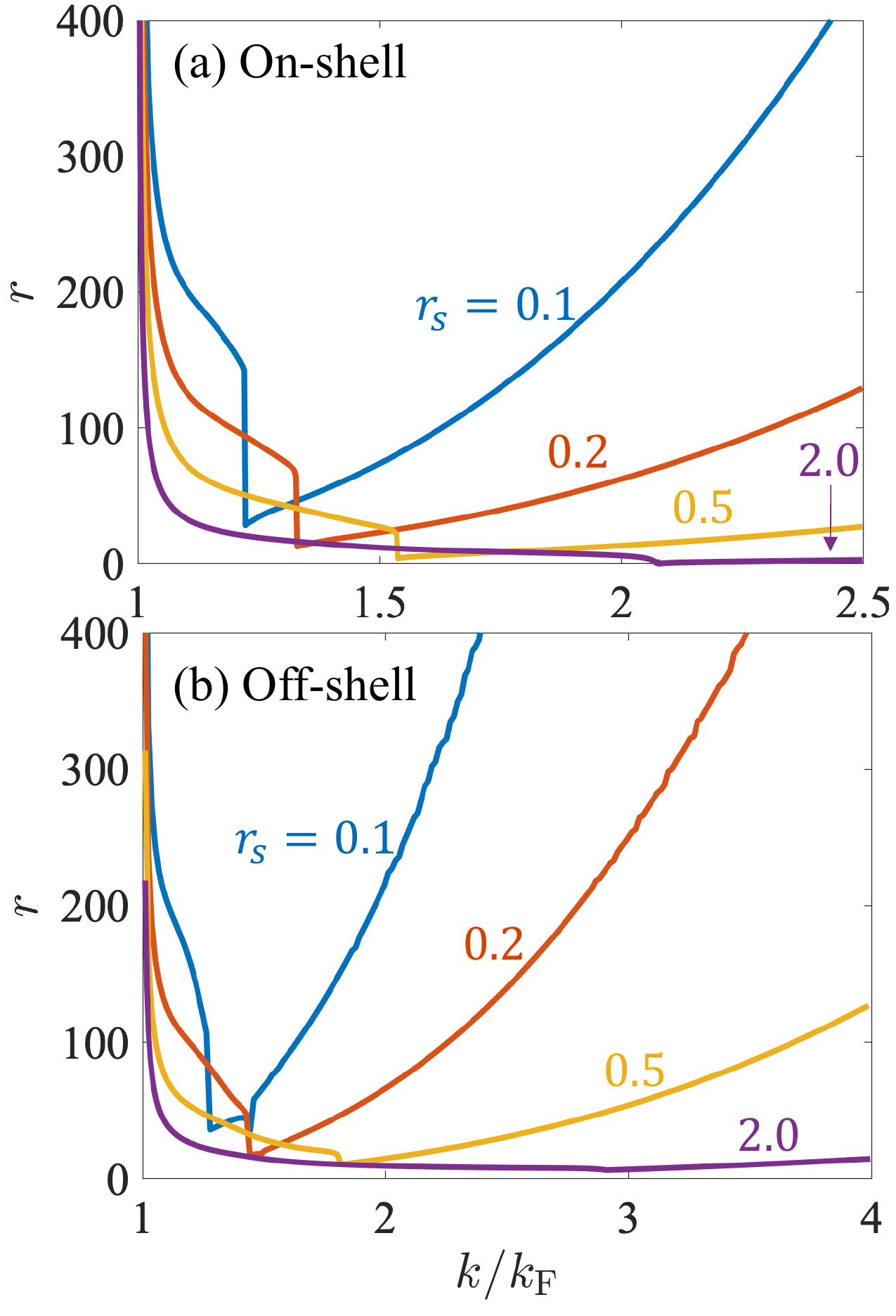}
  \caption{The ratio between the quasiparticle energy and the imaginary part of the self-energy within the (a) on-shell and (b) off-shell approximations for various values of $r_s=0.1$, $0.2$, $0.5$, and $2.0$.  }
  \label{fig:eqp_imag_ratio}
\end{figure}

 \begin{figure}[!htb]
  \centering
  \includegraphics[width=\linewidth]{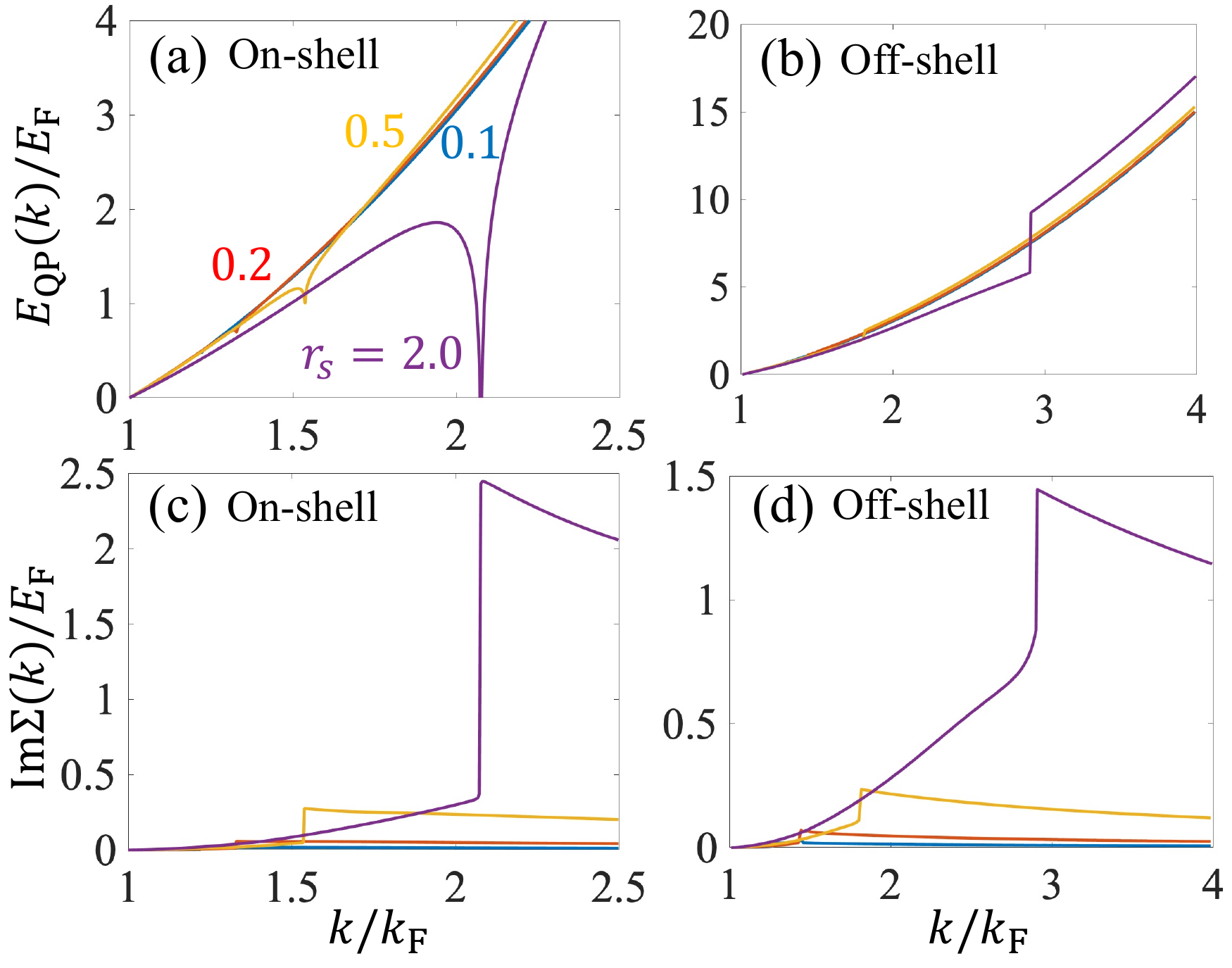}
  \caption{The calculated [(a) and (b)] quasiparticle energy and [(c) and (d)] imaginary part of the self energy obtained within the [(a) and (c)] on-shell and [(b) and (d)] off-shell approximations for various values of $r_s=0.1$, $0.2$, $0.5$, and $2.0$. }
  \label{fig:EQP_Imag}
\end{figure}

\begin{figure*}[!htb]
  \centering
  \includegraphics[width=\linewidth]{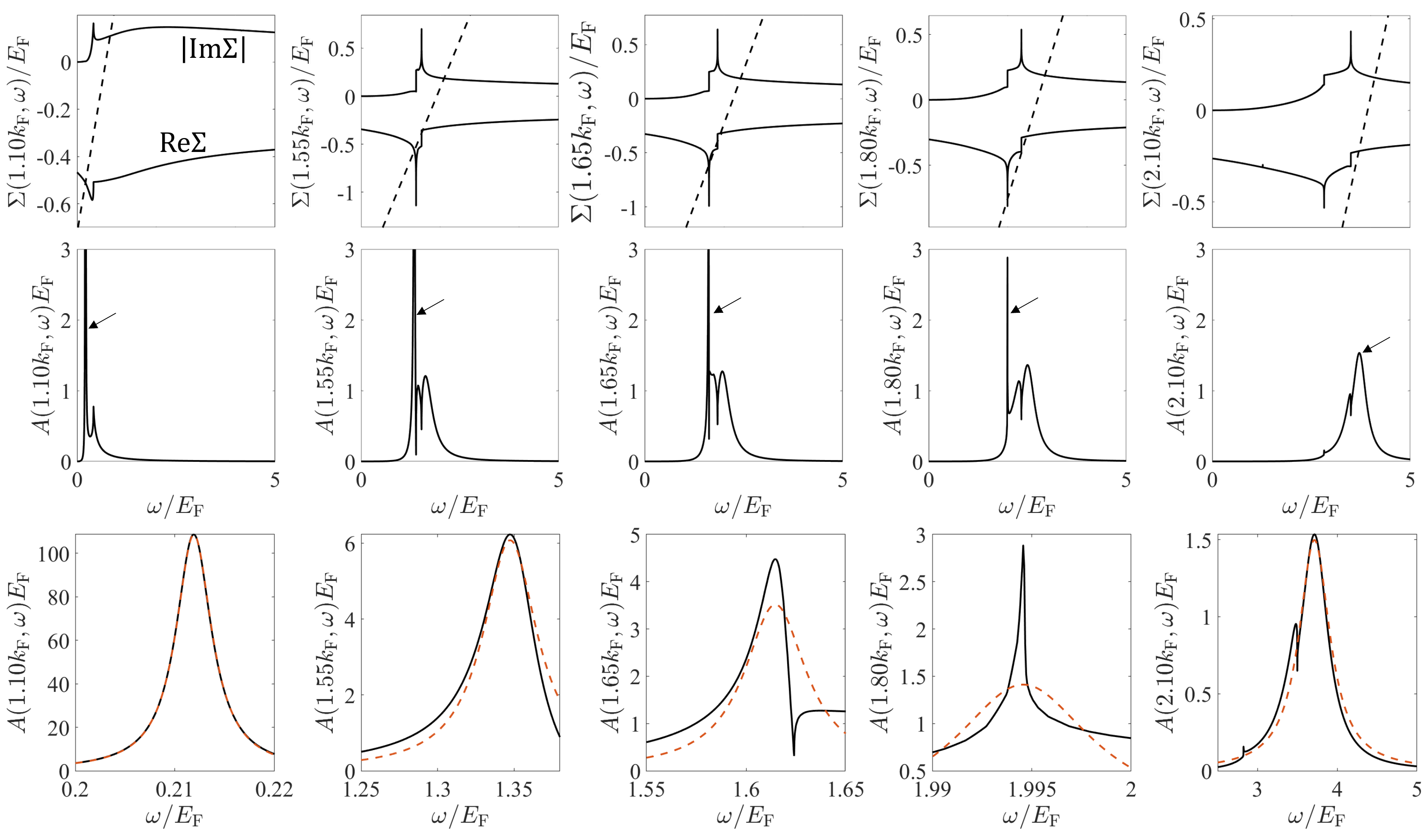}
  \caption{The upper five figures show the calculated real and imaginary part of self-energies for several fixed momenta $k/k_\mathrm{F}=1.1$, $1.55$, $1.65$, $1.75$ and $1.4$. For visual clarity, $|\mathrm{Im}\Sigma|$ is plotted instead of $\mathrm{Im}\Sigma$. The straight dashed lines are given by $\hbar\omega - \varepsilon_{\bm k} + E_\mathrm{F}$, whose intersection with $\mathrm{Re}\Sigma$ corresponds to the solutions of the Dyson's equation giving sharp peaks of the spectral functions plotted below. The figures in the third row show the zoom-in of the spectral peak along with the best fit curve by a Lorentzian distribution (dahsed red). Here we set $r_s=0.5$, and $\omega$ is measured from the interacting Fermi energy (see Appendix for results for different values of $r_s$).}
  \label{fig:spectral_functions}
\end{figure*}
\section{Validity of the quasiparticle picture} \label{sec:Validity_of_quasiparticle_picture}
The quasiparticle energy can be obtained by finding the poles of the interacting Green's function. From Eq.~(\ref{eq:interacting_green}), we can write the equation for the quasiparticle energy as 
\begin{equation}
    E_\mathrm{QP}(\bm k) - \varepsilon_{\bm k} + E_\mathrm{F} = \mathrm{Re}\Sigma[\bm k,E_\mathrm{QP}(\bm k)],
    \label{eq:Dyson}
\end{equation}
where $E_\mathrm{F}$ is the renormalized Fermi energy defined as
\begin{equation}
    E_\mathrm{F}=\varepsilon_{\bm k_\mathrm{F}} + \mathrm{Re}\Sigma(\bm{k}_\mathrm{F},0).
    \label{eq:interacting_mu}
\end{equation}
Equation~(\ref{eq:Dyson}) is called Dyson's equation, and should be solved self-consistently. 
The corresponding quasiparticle scattering rate, i.e., the inverse of the quasiparticle lifetime, is given by $\hbar/\tau_k=2\mathrm{Im}\Sigma[\bm k,E_\mathrm{QP}(\bm k)]$.

In some cases, there are multiple solutions to the Dyson's equation. To avoid ambiguity, in the following we choose $E_\mathrm{QP}$ to be the solution with the longest lifetime $\tau_k$.
Assuming that interaction strength is sufficiently weak ($r_s\ll1$), we can approximate the quasiparticle energy as the first iterative solution to the Dyson's equation
\begin{equation}
    E_\mathrm{QP}(\bm k) = \varepsilon_{\bm k}+\mathrm{Re}\Sigma(\bm k, \xi_{\bm k})
    \label{eq:onshell_dispersion}
\end{equation}
with the quasiparticle scattering rate given by $\hbar/\tau_k=2\mathrm{Im}\Sigma(\bm k, \xi_{\bm k})$. This is called the on-shell approximation, which is in good agreement with the off-shell self-consistent solution defined in Eq.~(\ref{eq:Dyson}) in the weak interaction limit ($r_s\ll1$). Obviously, the off-shell approximation provides the correct results if the exact self-energy is used. Within the leading-order dynamical RPA theory, which is used in this work, however, the off-shell approximation is considered worse in some cases since it mixes the perturbative orders solving Eq.~(\ref{eq:Dyson}) self-consistently \cite{DuBois1959, DuBois1959a, Lee1975, Ting1975, Vinter1975Correlation,  Zhang2005}. The question of which approximation yields more accurate results when the leading-order RPA self-energy is used is not obvious and remains open. In this work, we present results obtained using both approximations, and show that the same conclusion is reached regardless of which approximation is used as far as the validity of the quasiparticle picture is concerned although there are substantial quantitative differences between the results obtained from on-shell and off-shell solutions of Dyson's equation.

The standard way to verify the validity of the quasiparticle picture is to compare the quasiparticle energy with the scattering rate (the imaginary part of the self-energy): if Eq.~(\ref{eq:quasiparticle_condition}) is satisfied, the quasiparticle picture is valid. For example, the imaginary part of the self-energy must vanish faster than $E_\mathrm{QP}$ itself at low energies ($E_\mathrm{QP}\sim 0$) to ensure that the Fermi surface exists in the presence of interactions, and indeed this happens in 2D and 3D, but not in 1D \cite{Hu1993Many}. 
To quantify the robustness of the quasiparticle picture, it is useful to introduce the ratio of the quasiparticle energy to the imaginary part of the self-energy, defined as
\begin{equation}
    r=E_\mathrm{QP}(k)/\mathrm{Im}\Sigma(k,\xi_k)
\end{equation}
for (a) the on-shell 
and
\begin{equation}
    r=E_\mathrm{QP}(k)/\mathrm{Im}\Sigma[k,E_\mathrm{QP}(k)]
\end{equation}
for (b) the off-shell results. Here the quasiparticle energy $E_\mathrm{QP}$ is measured (as an excitation energy) from the interacting Fermi energy [i.e., $E_\mathrm{QP}(k_\mathrm{F})=0$]. For $k=k_\mathrm{F}$, the ratio $r$ is infinite by definition since the zero energy quasiparticles on the Fermi surface are stable with infinite lifetime as the imaginary part of the self-energy vanishes on the Fermi surface. Our goal is to obtain $r$ ($E_\mathrm{QP}$) as a function of $k/k_\mathrm{F}$ to see how far from the Fermi surface the quasiparticles remain stable. We can rewrite the quasiparticle validity condition Eq.~(\ref{eq:quasiparticle_condition}) as $r\gg 1$. 
Note that the better quasiparticles are defined, the larger the ratio our calculated $r$ should be. An equivalent statement is that larger the value of $r$ is compared with unity, more stable is the quasiparticle. Figure~\ref{fig:eqp_imag_ratio} presents the plot of $r$ as a function of the momentum $k$ up to high energies far above the Fermi energy ($k\sim 4 k_\mathrm{F}$). 
Near the Fermi surface, the imaginary part of the self-energy vanishes as $\mathrm{Im}\Sigma[k,E_\mathrm{QP}(k)]\sim E_\mathrm{QP}(k)^2\ln{E_\mathrm{QP}(k)}$ \cite{Zheng1996}, and thus $r$ diverges approaching the Fermi surface ($k\rightarrow k_F$) satisfying the quasiparticle condition given by Eq.~(\ref{eq:quasiparticle_condition}). As one moves away from the Fermi surface, the quasiparticle scattering rate increases with increasing $k$ because the volume of the phase space into which quasiparticles can decay increases. Thus, $r$ exhibits a monotonically decreasing behavior near the Fermi surface. The decreasing behavior continues until $k$ increases up to a certain critical wave vector $k_c$, where $r$ starts increasing with increasing $k$. Note that this is in sharp contrast to the common intuition that the scattering rate keeps on increasing as one moves away from the Fermi surface, and becomes eventually much larger than the quasiparticle energy (i.e., $r\rightarrow 0$ as $k\rightarrow \infty$), violating the quasiparticle condition Eq.~(\ref{eq:quasiparticle_condition}). 
For small $r_s=0.1$, where the theory is almost exact, it is obvious that the ratio $r$ even at large $k$ is comparable to that near the Fermi surface, showing that quasiparticles are well-defined in a wide range of energy, even very far from the Fermi surface, when the interaction strength is small. For a strongly interacting situation with a larger $r_s=2.0$, however, quasiparticles are not as robust as those for small $r_s$, since the ratio $r$ at large $k$ is much smaller than that near the Fermi surface. But even for $r_s=2$, the ratio $r>1$ for $k/k_\mathrm{F}= 4$ (and even for larger $k$ values) although the condition $r\gg 1$ is no longer satisfied for large $k\gg k_\mathrm{F}$.
Our results imply that the quasiparticle picture remains robust at high energies than has been thought and the quasiparticle robustness at high energies is sensitive to the interaction strength. With increasing interaction strength (i.e., $r_s$) the quasiparticle stability characterized by the ratio $r$ becomes weaker, but $r>1$ minimal condition for the validity of the quasiparticle picture applies even for large $r_s$ and large $k/k_\mathrm{F}$. 
Our qualitative conclusion on the robust validity of the quasiparticle picture applies equally well to both on-shell and off-shell approximations, as is obvious from the two panels in Fig.~\ref{fig:eqp_imag_ratio}, although there are quantitative differences between the two approximations. Our conclusion based on the on-shell approximation is consistent with the recent analytical results in Ref. \cite{Sarma2021know} where the self-energy was calculated in a series expansion in $r_s$ and $E/E_\mathrm{F}$ using the on-shell approximation.

A noteworthy feature of Fig.~\ref{fig:eqp_imag_ratio} is the sharp minimum in the ratio $r$ at the intermediate $k=k_c$, where the 2D quasiparticle picture is the least stable (or most fragile) with the quasiparticles being more stable for both $k<k_c$ and $k>k_c$. Around this critical momentum $k_c$, the 2D quasiparticles are fragile (or the least stable) with $k_c$ being relatively small: $k_c \sim 1.5 k_\mathrm{F}$ (i.e. meaning, not much larger than $k_\mathrm{F}$). The existence of this `critical' momentum $k_c$ characterizing maximal 2D quasiparticle fragility would be further analyzed later in this article.

In Fig.~\ref{fig:EQP_Imag}, we show for completeness the actual 2D quasiparticle energy and the quasiparticle decay rate or broadening for both on-shell and off-shell approximations leading to the calculated $r$ in Fig.~\ref{fig:eqp_imag_ratio}. The existence of the critical $k_c$ is apparent in these results already with a local maxima in the decay rate in the quasiparticles.

\begin{figure*}[!htb]
  \centering
  \includegraphics[width=\linewidth]{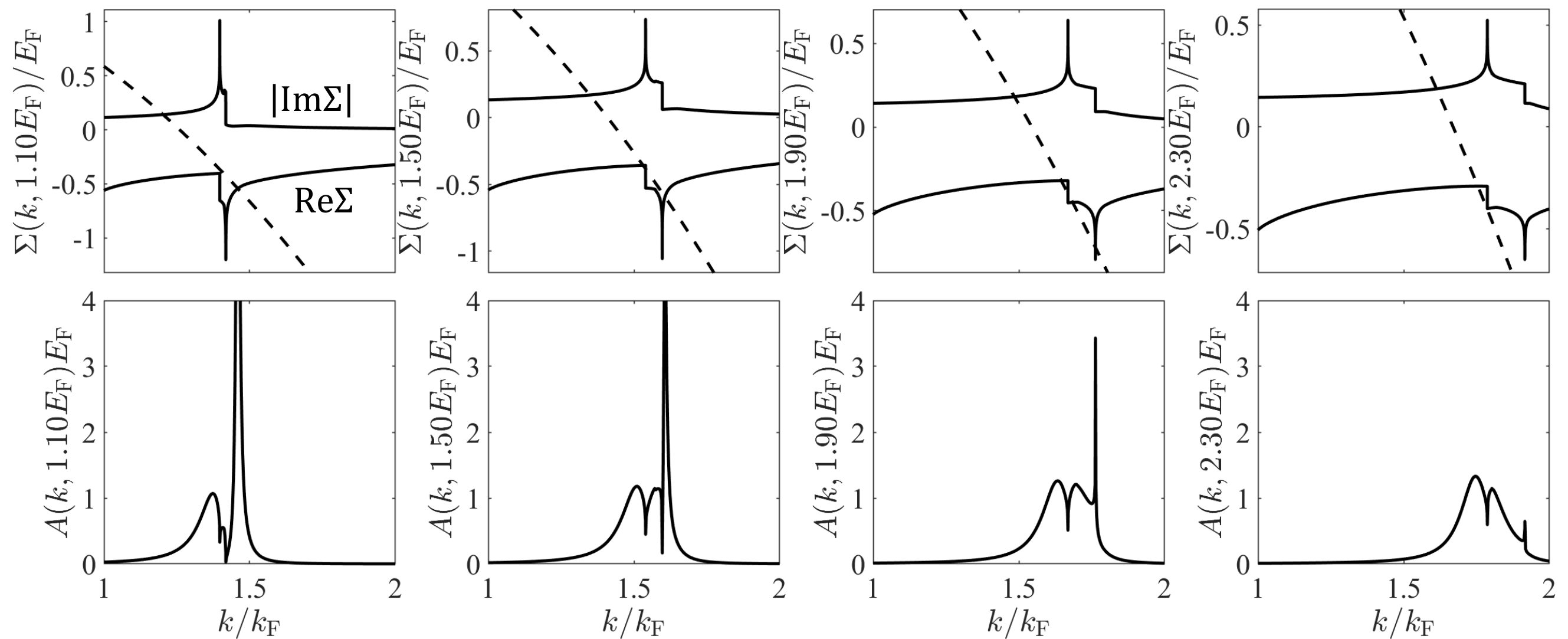}
  \caption{Plot of self-energy (first row) and spectral function (second row) as a function of momentum for several values of energies $\hbar\omega=1.1E_\mathrm{F}$, $1.5E_\mathrm{F}$, $1.9E_\mathrm{F}$ and $2.3E_\mathrm{F}$. For visual clarity, $\left|\mathrm{Im}\Sigma\right|$ is plotted instead of $\mathrm{Im}\Sigma$. The straight dashed lines are given by $\hbar\omega - \varepsilon_{\bm k} + E_\mathrm{F}$, whose intersection with $\mathrm{Re}\Sigma$ corresponds to the solutions of the Dyson's equation giving sharp peaks of the spectral functions. Here we set $r_s=0.5$.}
  \label{fig:spectral_function_momentum}
\end{figure*}

\section{spectral function} \label{sec:spectral_function}
Figure~\ref{fig:spectral_functions} presents calculated self-energies (first row) and the corresponding spectral functions (second row). The intersection between the dashed straight-line and the real part of the self-energy represents the self-consistent off-shell solutions to the Dyson's equation. 
It is well known that the spectral function appears as a $\delta$-function at the Fermi surface ($k=k_F$) since the imaginary part of the self-energy vanishes as $k\rightarrow k_\mathrm{F}$ behaving as $\mathrm{Im}\Sigma\sim \omega^2 \ln{\omega}$ \cite{Zheng1996}. This establishes the perturbative preservation of the interacting 2D Fermi surface as postulated in the Landau Fermi liquid theory. As one moves away from the Fermi surface, the quasiparticle decays through the creation of electron-hole pairs, and the spectral function evolves into a single broadened Lorentzian type peak with a finite linewidth as shown in the result for $k=1.1k_\mathrm{F}$. Whether the spectral shape is precisely Lorentzian or not depends crucially on the detailed energy dependence of the self-energy.
The spectral peak continues to broaden with increasing $k$ due to the increasing $\mathrm{Im}\Sigma$. As we approach the critical wave vector $k_c=1.8k_\mathrm{F}$, however, the spectral peak corresponding to the solution closest to $\omega=0$ indicated by black arrows in the figure becomes sharper despite the increasing imaginary part of the self-energy with increasing $k$. Note that there are multiple solutions to the Dyson's equation for the results around the critical wave vector $k_c$ ($k=1.55k_\mathrm{F}$, $1.65k_\mathrm{F}$, and $1.8k_\mathrm{F}$), giving rise to overlapped damped-peaks in the spectral function. These multiple solutions of Dyson's equation already indicate a nominal failure of the quasiparticle picture around $k_c$ since a quasiparticle should be a unique well-defined spectral peak. This spectral structure around $k_c$ is not obviously quasiparticle-like since it is non-Lorentzian and very broad due to a large imaginary part of the self-energy. Thus, we ignore them in the following discussion, and only refer to the solution (peak) with the smallest imaginary part of the self-energy (i.e., the lowest energy solution or the one closest to the Fermi surface $\omega=0$) as quasiparticle solution (peak).
The figures in the third row zoom on the quasiparticle peaks indicated by black arrows in the second row. Note that the quasiparticle peak deviates more from the Lorentzian shape as $k\rightarrow k_c$. In particular, at the critical wave vector ($k=1.8k_\mathrm{F}$) the spectral function exhibits a completely arbitrary shape that does not fit to the Lorentzian curve at all. Thus, for $k\sim k_c$, the 2D quasiparticle is fragile, as inferred already in Sec.~\ref{sec:theory} based on the calculated $r$ values (Fig.~\ref{fig:eqp_imag_ratio}).
For $k>k_c$, the spectral function eventually recovers the Lorentzian shape, but with a large spectral width-to-height ratio compared to those near the Fermi surface, meaning that the quasiparticle picture is less robust than near the Fermi surface. Thus, the quasiparticles are generally more stable for $k<k_c$ than for $k>k_c$ since their spectral functions are typically much broader above $k_c$. The quasiparticle picture at large $k>k_c$ will be discussed further in the following sections.

Such an anomalous behavior of the quasiparticle peak also occurs in the momentum dependence of the spectral function at fixed energy.
Figure~\ref{fig:spectral_function_momentum} shows the plot of the self-energy and spectral function as a function of momentum. Note that the momentum-dependent spectral function shows a similar behavior as the energy spectral function plotted in Fig.~\ref{fig:spectral_functions}: The spectral function near the Fermi surface exhibits a typical Lorentzian quasiparticle peak with a small broadening. As one moves away from the Fermi surface with increasing $\omega$ approaching the critical energy ($E_c=1.9E_\mathrm{F})$, the broadening of the peak rapidly decreases, becoming much sharper than the one near the Fermi surface ($E=1.1E_\mathrm{F}$), but the spectral shape becomes highly non-Lorentzian with considerable incoherent spectral weight. Above the critical energy ($E=2.3E_\mathrm{F}$), the spectral peak recovers its Lorentzian shape with a very large broadening, similarly to the energy-dependent spectral function discussed previously.

Our calculated results for the self-energy and spectral function show that quasiparticles are fragile in a small discrete region around the critical wave vector $k_c$, where the spectral function is highly non-Lorentzian and mostly incoherent. In the next section, we further investigate the quasiparticle fragility, revealing its origin to be the coupling of quasiparticles with the plasmon collective excitations, which become important at higher momenta.
We mention that our calculated spectral function can be directly explored experimentally by carrying out either energy-resolved spectroscopy at fixed momentum or by momentum resolved spectroscopy at fixed energy. It is also, in principle, possible to directly measure the momentum and energy dependent spectral function by using tunneling spectroscopy.

\section{renormalization factor} \label{sec:renormalization_factor}
We can decompose the spectral function into the coherent quasiparticle $A_\mathrm{QP}$ and incoherent $A_\mathrm{inc}$ parts:
\begin{equation}
    A(k,\omega)=A_\mathrm{QP}(k,\omega)+A_\mathrm{inc}(k,\omega)
\end{equation}
where $A_\mathrm{QP}(k,\omega) = -\frac{1}{\pi} \mathrm{Im} G_\mathrm{QP}(k,\omega)$ with
\begin{equation}
    G_\mathrm{QP}(k,\omega)=\frac{Z_k}{\hbar\omega-E_\mathrm{QP}(k)-iZ_k\mathrm{Im}\Sigma(k, \omega)}     
\end{equation}
being the singple particle Green's function for the quasiparticle. $A_\mathrm{inc}(k,\omega)=A(k,\omega)-A_\mathrm{QP}(k,\omega)$ absorbs all the incoherent contribution to the spectral function. $Z_k$ is the renormalization factor defined as \cite{Rice1965, fetter2012quantum, mahan2000many}
\begin{equation}
    Z_k=\left(\left.1- \frac{\partial \mathrm{Re} \Sigma(k,\omega)}{\hbar\partial \omega}\right\rvert_{\hbar\omega=E_\mathrm{QP}(k)}\right)^{-1}.
\end{equation}
The renormalization factor $Z_k$ determines the transfer of interaction-induced spectral weight from the coherent quasiparticle to the incoherent non-quasiparticle part, and thus should be a positive number between 0 and 1 if the quasiparticle picture is valid. $Z_k$ can also be equivalently thought to be the effective wavefunction overlap between the interacting quasiparticle and the noninteracting particle as the interaction is turned on adiabatically, again implying that the quasiparticle picture applies only when $0<Z<1$, with $Z$ providing the ``renormalization'' of the noninteracting electron into becoming the interacting ``quasiparticle''.

\begin{figure}[!htb]
  \centering
  \includegraphics[width=\linewidth]{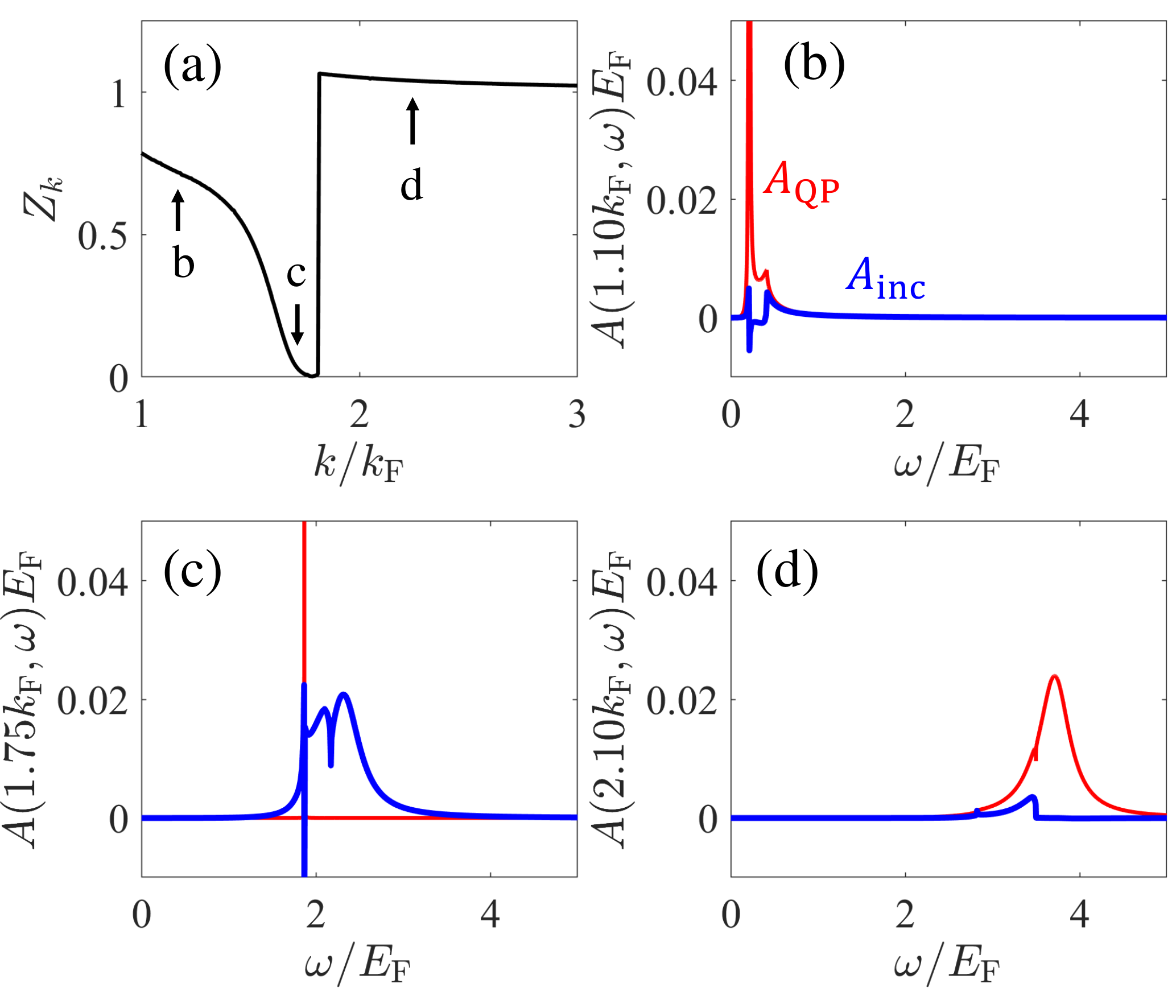}
  \caption{(a) Plot of the renormalization factor $Z_k$ as a function of momentum $k$.(b), (c), (d) spectral function decomposed into the coherent part $A_\mathrm{QP}$ (red) and incoherent part $A_\mathrm{inc}$ (blue) at a wave vector (b) below $k_c$, (c) near $k_c$, and (d) above $k_c$. For convenience, we indicate the corresponding value of $Z_k$ with arrows in (a). Here $r_s=0.5$ is used.}
  \label{fig:Zk_AQP_Ainc}
\end{figure}

Figure~\ref{fig:Zk_AQP_Ainc} shows the plot of $Z_k$ and the spectral function decomposition into the coherent (blue) and incoherent (red) parts. Near the Fermi surface, $Z_k$ decreases as one moves away from the Fermi surface with increasing $k$. As $k\rightarrow k_c$, $Z_k$ rapidly drops down to almost zero, which is not surprising given that the spectral peak shape in this regime loses the coherent Lorentzian structure as shown in the previous section. It is worth noting that such an anomalous behavior of $Z_k\rightarrow 0$ also occurs in one-dimensional electron system where even the low energy excitations have a collective nature \cite{Tomonaga1950, Luttinger1963} except that in 1D this happens at all momenta destroying the Fermi surface, thus converting the 1D interacting system into a Luttinger liquid with no Fermi surface. Since $Z_k$ represents the relative weight of the quasiparticle part to the incoherent background, a small value of $Z_k$ implies that the incoherent background completely dominates over the quasiparticle coherent structure. The quasiparticle picture is then fragile. This can be more clearly seen in Figs.~\ref{fig:Zk_AQP_Ainc}(b) and \ref{fig:Zk_AQP_Ainc}(c): the spectral peak near the Fermi surface [Fig.~\ref{fig:Zk_AQP_Ainc}(b)] consists only of the coherent quasiparticle part of the spectral function, i.e., $A_\mathrm{QP}$, with the small background contributed by the incoherent part $A_\mathrm{QP}$. For $k\sim k_c$, however, the spectral peak is contributed by both coherent and incoherent components and the spectral function is highly non-Lorentzian with no obvious quasiparticle peak anymore although there is a spectral peak. This indicates that even though the spectral peak is sharp, much of it comprises of the incoherent part being of non-quasiparticle nature, and thus the quasiparticle picture in the regime around the critical wave vector $k_c$ is inapplicable or in other words, the 2D quasiparticle is fragile around $k_c$. Even though here we present the spectral results only for $r_s=0.5$, the same conclusion about fragility around a critical momentum applies regardless of the value of $r_s$, even for very small weakly interacting $r_s\ll1$ where our leading-order dynamical approximation is exact (see Fig.~\ref{fig:spectral_Aqc_Ainc_Appendix} in Appendix).
As we go above $k_c$, $Z_k$ suddenly increases up to the value larger than unity (approximately $1.02$), and gradually decreases with increasing $k$, approaching unity from above as k approaches infinity. This anomalous behavior of $Z_k$ being larger than unity occurs due to the positive slope of the real part of the self-energy [see the inset of Fig. \ref{fig:Zk_AQP_Ainc}(d)]. Note that $Z_k>1$ is pathological, and is not consistent with the physical interpretation of the renormalization factor in the quasiparticle theory (e.g., the wavefunction overlap between the noninteracting and interacting states), and thus is a sign for the breakdown of the quasiparticle picture. 
In strongly correlated lattice models, where the physics is very different from our continuum interacting electron liquid jellium model, the finding of $Z_k>1$ is sometimes associated with the complete breakdown of the Fermi liquid behavior (and hence, the quasiparticle picture), as for example, in the 2-channel Kondo lattice problem \cite{Cox1996}, but whether our finding of $Z_k>1$ for $k>k_c$ signals such a breakdown of the Fermi liquid theory is doubtful.
The fact that our calculated $Z_k$ eventually approaches unity at sufficiently high energies indicates that the very slightly larger than unity ($Z\sim 1.02$) value of the renormalization factor for the interacting 2DEG in our theory for $k>k_c$ is just an indirect signature of the existence of the critical momentum $k_c$ where the Fermi liquid description breaks down at a set of measure zero (i.e., just at one momentum away from $k_\mathrm{F}$). We note that $Z_k$ approaching unity at very high energies is expected because at extremely high energies, the interacting system should behave as non-interacting.

\begin{figure}[!htb]
  \centering
  \includegraphics[width=\linewidth]{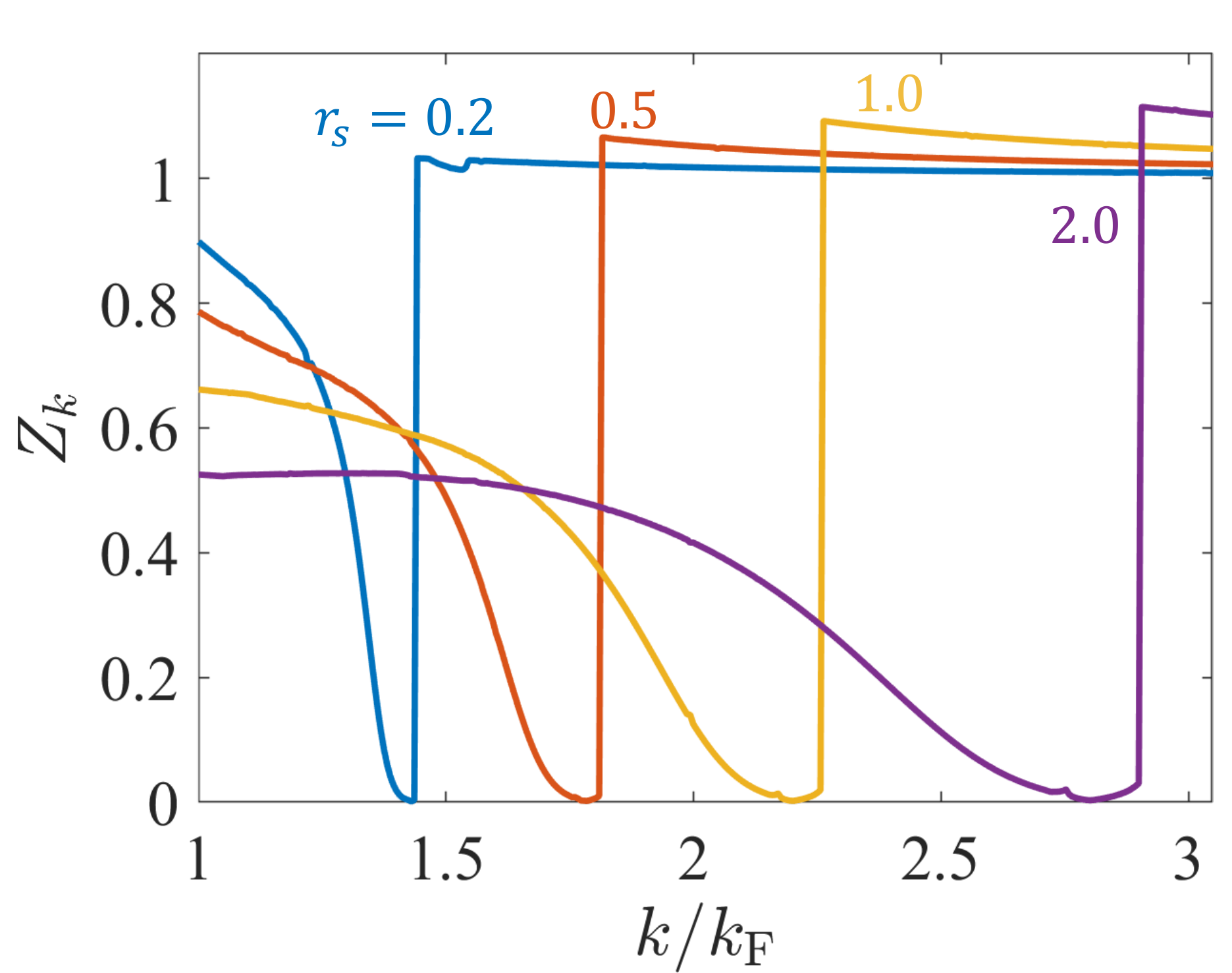}
  \caption{Renormalization factor $Z_k$ plotted as a function of the momentum $k$ for several values of $r_s=0.2$, $0.5$, $1.0$, and $2.0$. }
  \label{fig:Zk_various_rs}
\end{figure}

Figure~\ref{fig:Zk_various_rs} shows the plot of $Z_k$ for various values of $r_s$. $Z_k$ at the Fermi surface ($k=k_\mathrm{F}$) represents the Fermi surface discontinuity in the interacting system through the relation $Z_{k_\mathrm{F}}=n_{k_\mathrm{F}-\epsilon}-n_{k_\mathrm{F}+\epsilon}$, which decreases with increasing interaction strength ($r_s$), implying that the Fermi surface is suppressed as interaction strength increases. For all values of $r_s$, $Z_k$ also decreases with increasing $k$ up to the critical wave vector $k_c$. However, it is important to note that the range of $k$ below $k_c$ where $Z_k$ is smaller than a certain small value (e.g., $Z_k=0.1$), is much broader for large $r_s$ (e.g., $r_s=2.0$) than that for small $r_s$ (e.g., $r_s=0.2$). This shows that the quasiparticle picture is much more fragile in the strongly interacting regime, as expected.
For all values of $r_s$, $Z_k$ jumps to a value larger than unity just above $k_c$ as is obvious from $Z_k$ becoming close to unity above $k_c$. Note that the maximum of $Z_k$, which occurs at $k=k_c$, is larger as $r_s$ increases. This shows that the quasiparticle behavior at large $k$ above the critical wave vector $k_c$ becomes more fragile as the interaction strength increases. It is also important to note that $Z_k$ decreases with increasing $k$ for $k>k_c$ for all values of $r_s$. These results imply that the quasiparticle picture becomes more robust with increasing $k$ regardless of the interaction strength.
This is also consistent with the interacting system having a $Z_k$ approaching unity for very large $k$, which implies almost free-electron type behavior at very high momenta. These results are consistent with the 2D quasiparticles being fragile for all $r_s$ around $k\sim k_c$, and being reasonably stable at all other momenta away from $k_c$. In particular, Fig.~\ref{fig:Zk_various_rs} for $Z_k$ as a function of $k$ suggests that the quasiparticles become continuously less stable as k increases from $k_\mathrm{F}$ to $k_c$, reaching the least robust character at $k=k_c$, where $Z_k$ almost vanishes, but for $k>k_c$, the quasiparticles are again very robust, and are essentially like free particles with $Z_k \sim, 1$. Thus, the fragile quasiparticles reside at $k=k_c$, and just below it. This is also consistent with the calculated $r$ given in Fig.~\ref{fig:eqp_imag_ratio}.
Last, we emphasize that $Z_k$ vanishes at the critical wave vector $k_c$ regardless of the value of $r_s$ even in the weakly interacting limit ($r_s=0.2$) where our RPA theory is exact. This indicates that the vanishing $Z_k$ is not just a mere artifact of the leading order approximation even though including higher order terms may slightly change the results (e.g., the precise value of $k_\mathrm{c}$) quantitatively.

\begin{figure}[!htb]
  \centering
  \includegraphics[width=\linewidth]{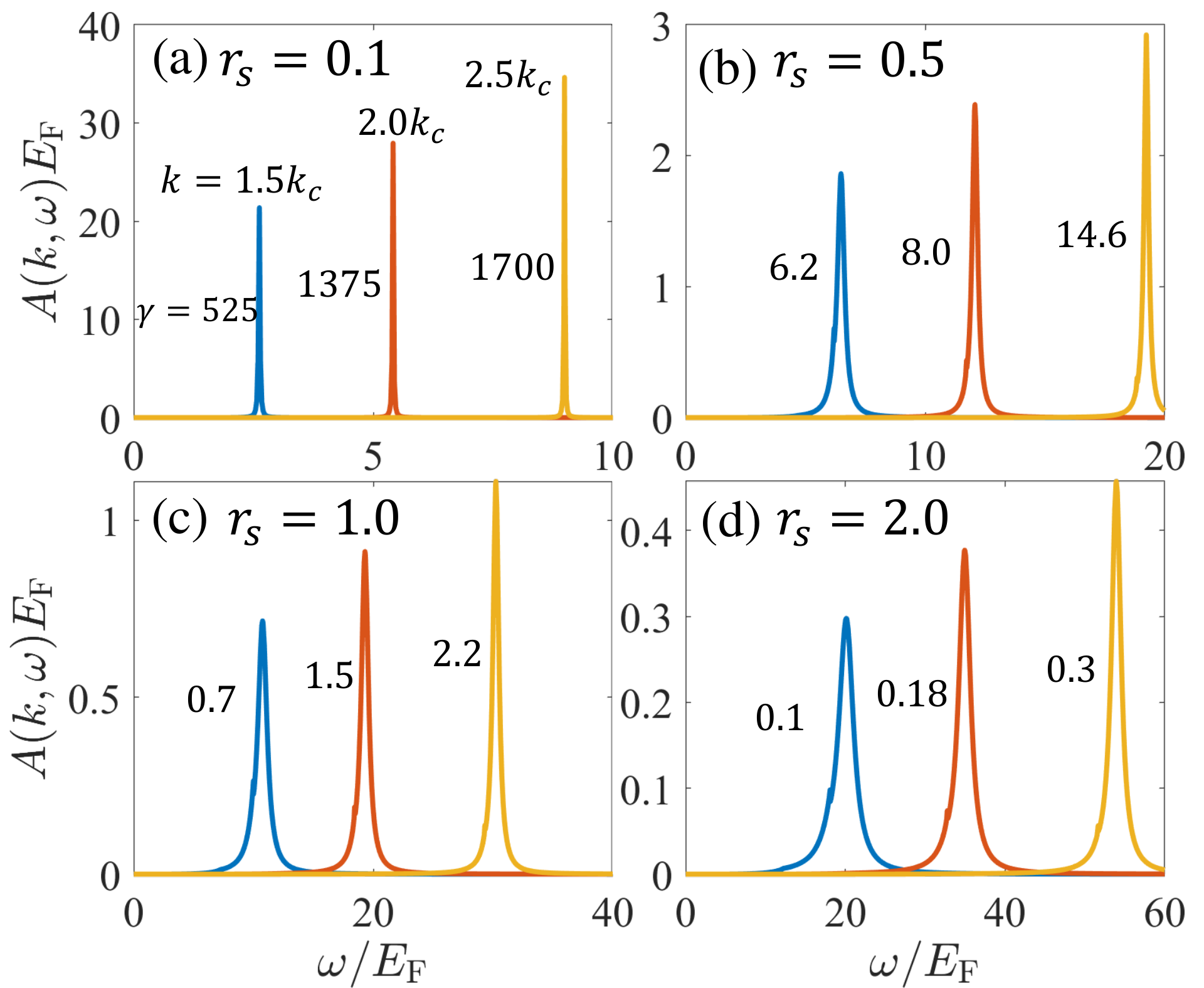}
  \caption{Spectral functions $A(k,\omega)$ at large momenta $k=1.5k_c$, $2.0k_c$, and $2.5k_c$ are plotted together as a function of the energy $\omega$ for various values of $r_s=0.1$, $0.2$, $0.5$, and $2.0$. $\gamma$ (as labeled on each curve of the spectral function in each panel) indicates the dimensionless ratio of the spectral height to the spectral width.}
  \label{fig:spectral_function_at_large_k}
\end{figure}

To further investigate quasiparticle fragility versus stability, we plot in Fig.~\ref{fig:spectral_function_at_large_k} the calculated spectral functions at various values of $k>k_c$ for different values of $r_s$. Here the broadening of the spectral peak is quantified by $\gamma$, which is the ratio between the height and width of the spectral peak, defined as
\begin{equation}
 \gamma=\frac{A(k, E_\mathrm{QP})E_\mathrm{F}}{\Delta \hbar\omega/E_\mathrm{F}}
\end{equation}
where $\Delta\omega$ is the width of the spectral peak defined through $A(k,E_\mathrm{QP}\pm \Delta\hbar\omega) = A(k,E_\mathrm{QP})/2$ (Changing the factor of 2 some other number does not change any conclusion.). 
Dimensionalizing by $E_F$ converts both the height and the width of the spectral function into natural dimensionless numbers, enabling a direct comparison between the height and the width.
Two remarks are in order. First, all the spectral peaks have a Lorentzian shape for all values of $r_s$ despite $Z_k>1$ and the broadening of the spectral peak increases as the interaction strength $r_s$ increases. Second, with increasing $k$, the spectral peak becomes sharper, indicating that the quasiparticle picture becomes more robust with increasing $k$. This is consistent with Fig.~\ref{fig:eqp_imag_ratio} where $r$ increases with increasing $k$ in the regime $k>k_c$, and also with the calculated renormalization factor $Z_k$ in Fig.~\ref{fig:Zk_various_rs}.
It should be noted that even though $\gamma$ increases with increasing $k$, the quasiparticle peak does not become infinitely sharp even at a reasonably large $k>k_c$. This is because there is always available phase space into which quasiparticles can decay unless they are located exactly at the Fermi surface (where the spectral function has a perfectly sharp $\delta$-function quasiparticle peak), and thus the spectral peak always has a finite width corresponding to a finite lifetime.
Note that for large $r_s$, $\gamma$ is very small (less than unity), and thus the quasiparticle peak is very broad. 
Whether such broad peaks can be construed to represent quasiparticles or not is basically a matter of semantics as they satisfy the requirement of $E_\mathrm{QP}$ being larger than the imaginary part of the self-energy, but at the same time are very broad with short quasiparticle lifetimes.
There is no sharp criterion distinguishing between quasiparticles and non-quasiparticles based on the broadening of the spectral peak compared with the peak height.
However, for small $r_s=0.1$, $\gamma$ is of the order of $10^3$ at large $k$, which is comparable to $\gamma$ of the well-defined quasiparticle peak at low energies near the Fermi surface. Thus, our results imply that for small $r_s$, the quasiparticle concept is valid even at high energies far away from the Fermi surface, and this conclusion most likely remains valid for large $r_s$ also although the quasiparticles are less sharply defined in this case (and additionally, our theory is less accurate)

\begin{figure}[!htb]
  \centering
  \includegraphics[width=\linewidth]{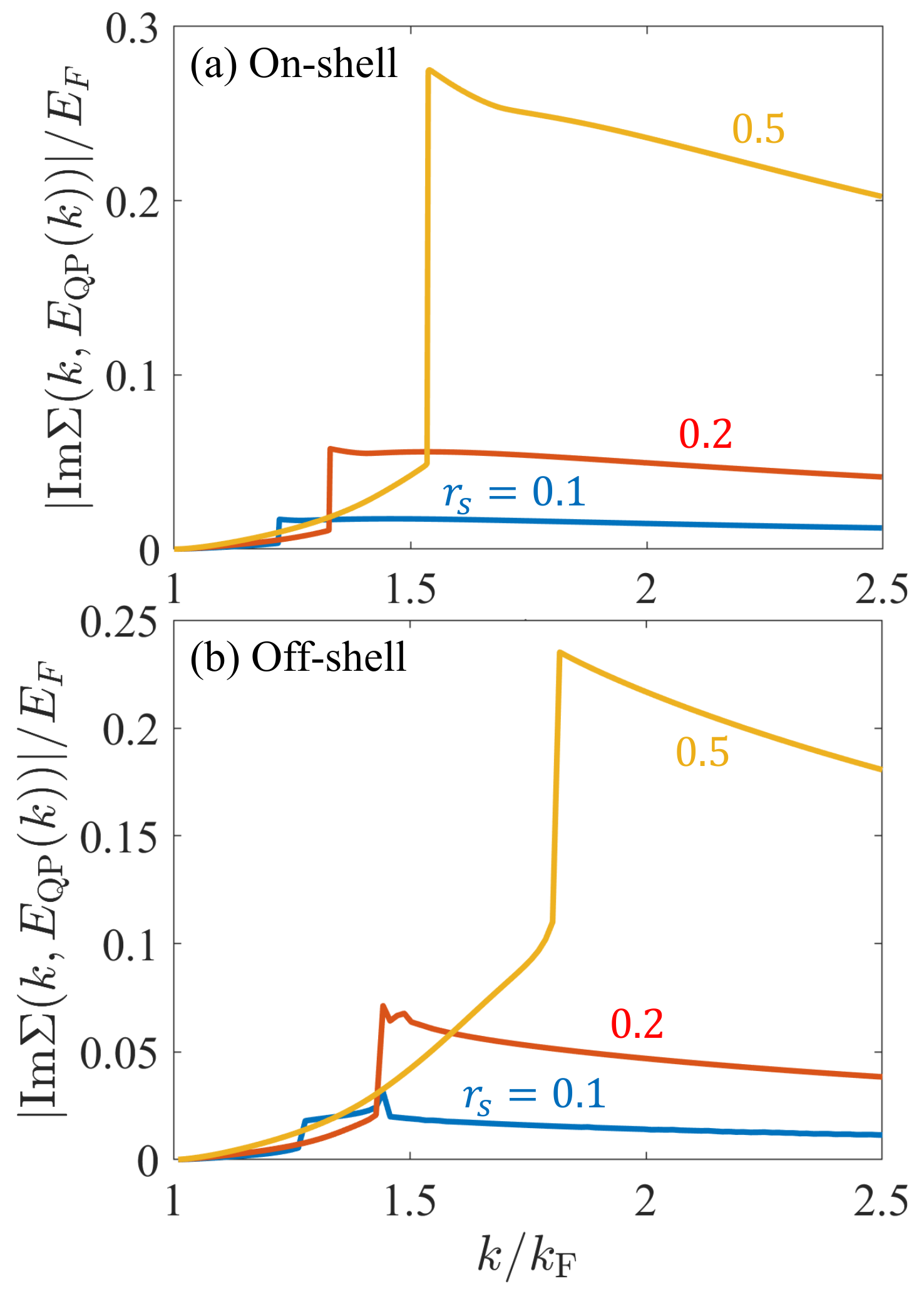}
  \caption{The imaginary part of the self-energy within the (a) on-shell and (b) off-shell approximations for various values of $r_s=0.1$, $0.2$ and $0.5$.  }
  \label{fig:imaginary_part_large_k}
\end{figure}

Figure~\ref{fig:imaginary_part_large_k} shows the imaginary part of the self-energy at large $k>k_c$ within both on-shell and off-shell approximations. Near the Fermi surface, the imaginary part of the self-energy manifests the expected quasiparticle behavior, $\mathrm{Im}\Sigma[k,E_\mathrm{QP}(k)]\sim E_\mathrm{QP}(k)^2\ln{E_\mathrm{QP}(k)}$. As we move away from the Fermi surface, the imaginary part of the self-energy increases due to the decay process through electron-hole emission. At the critical wave vector $k_c$, the imaginary part of the self-energy exhibits an abrupt jump. Within the on-shell approximation, it is well understood that this abrupt jump occurs because an additional decay channel via plasmon emissions is turned on \cite{Giuliani1982} as energy-momentum conservation creates a sharp threshold for plasmon emission at $k=k_c$. This shows that the fragility of the quasiparticle picture around $k_c$ arises from the collective plasmon mode. For large $k>k_c$, the imaginary part of the self-energy decreases with increasing $k$. This explains why the quasiparticle peak is better defined in the interacting spectral function as $k$ increases.

\begin{figure}[!htb]
  \centering
  \includegraphics[width=\linewidth]{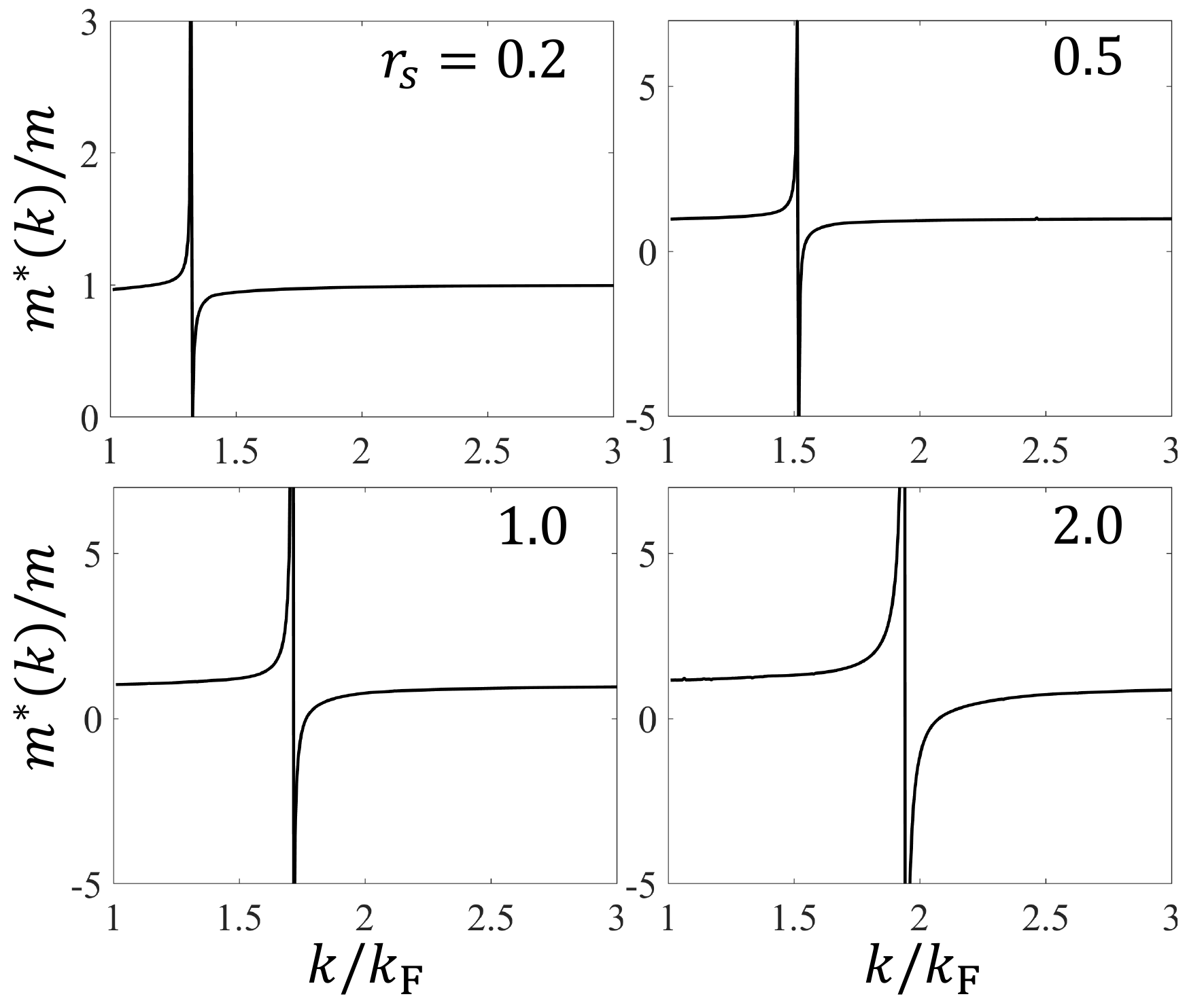}
  \caption{Calculated on-shell effective mass $m^*$ as a function of the momentum $k/k_\mathrm{F}$ for various values of $r_s=0.2$, $0.5$, $1.0$ and $2.0$. }
  \label{fig:effective_mass}
\end{figure}

\begin{figure*}[!htb]
  \centering
  \includegraphics[width=\linewidth]{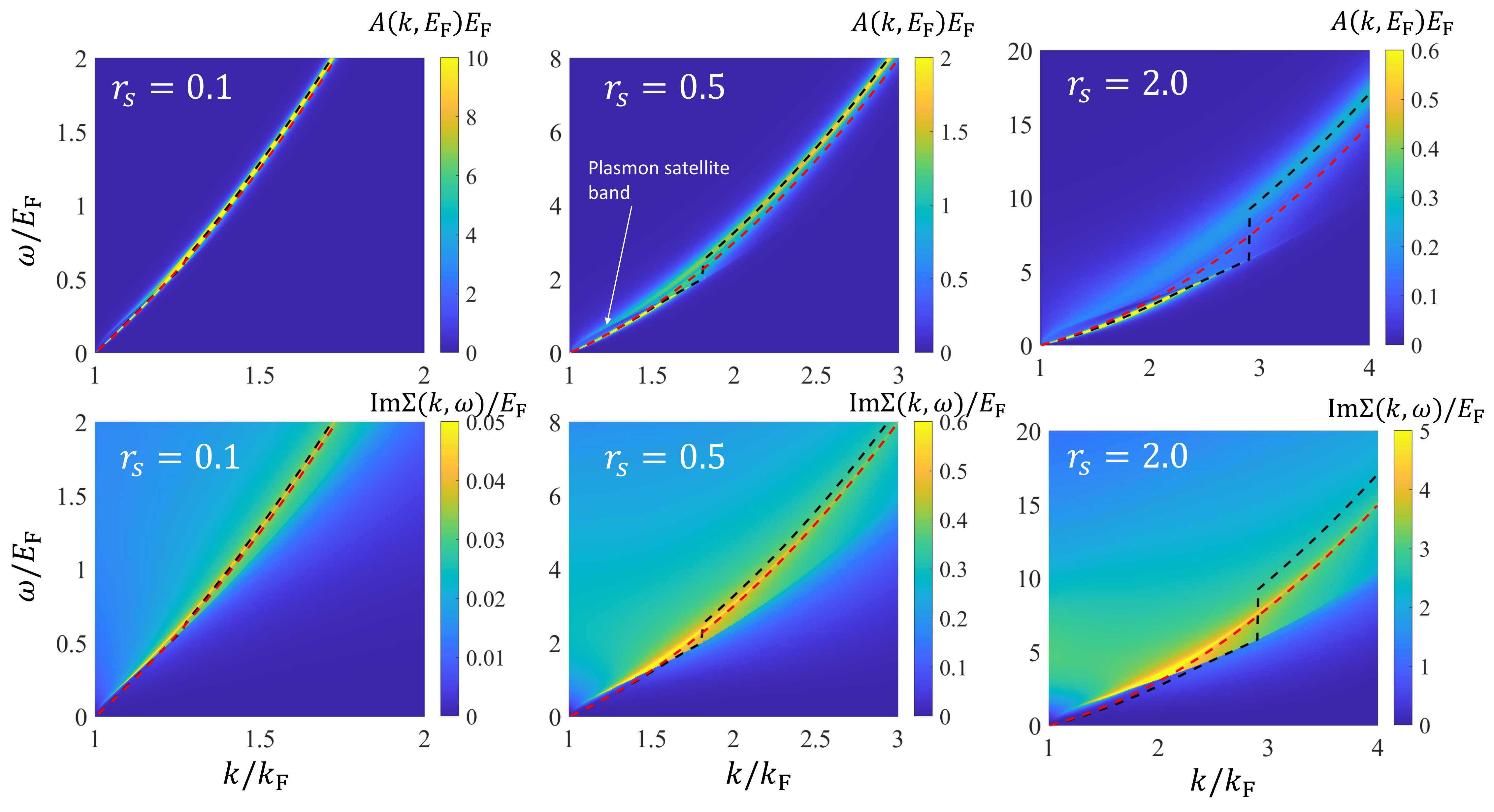}
  \caption{Two-dimensional plot of the spectral function (top) and the imaginary part of the self-energy (bottom) for various values of $r_s$. The black and red dashed line represents the renormalized quasiparticle band dispersion obtained within the off-shell approximation and the bare noninteracting energy dispersion given by $E(k)/E_\mathrm{F}=(k/k_F)^2-1$, respectively.  The white arrow indicates the plasmon satellite band, which disperses as $\omega\sim\sqrt{k-k_\mathrm{F}}$ at small wave vectors and merges into the quasiparticle energy band at the critical wave vector $k_c$. Here all the energies are measured from the interacting Fermi energy. }
  \label{fig:density_plot}
\end{figure*}

The renormalization factor $Z_k$ is defined only in the off-shell approximation. The equivalent quasiparticle quantity defined in the context of the on-shell approximation is the momentum-dependent effective mass $m^*(k)$. Using the Eq.~(\ref{eq:onshell_dispersion}), it is easy to obtain the effective mass:
\begin{equation}
    m^*( k)=\left\{
     1 + \frac{m}{\hbar^2k}\frac{\partial\mathrm{Re}[\Sigma( k,\xi_{ k})] }{\partial k} \right\}^{-1}.
\label{eq:effective_mass}
\end{equation}
In Fig.~\ref{fig:effective_mass}, we plot the renormalized effective mass for various values of $r_s$. The effective mass is well defined near the Fermi surface. The effective mass rapidly grows, eventually diverging as approaching the critical wave vector $k_c$, and then suddenly drops to a negative value just above $k_c$. Such an unphysical behavior of the effective mass is not consistent with the quasiparticle theory, thus implying that the quasiparticle picture is suspect around the critical wave vector $k_c$. We mention that such a divergence of the effective mass has earlier been reported in strongly interacting electron systems at large $r_s$ and is considered a sign of the breakdown of the Fermi liquid quasiparticle theory \cite{Zhang2005}. As we move further away from the critical wave vector $k_c$ with increasing $k$, the effective mass recovers to a physical positive value from below, exhibiting a typical quasiparticle behavior for $k>k_c$. Note the range of $k$ where the renormalized effective mass has a non-physical behavior is wider for larger $r_s$, and thus the interaction strength strongly affects the range of validity of the Fermi liquid theory. We emphasize that all these results based on the on-shell approximation are consistent with the off-shell results discussed above in the sense that both show that the quasiparticles become fragile with a potential breakdown of the Fermi liquid picture for $k\sim k_c$.

So far, we have seen that the quasiparticle picture is fragile in a regime around $k_c$. Interestingly, this happens for all values of $r_s$ regardless of the strength of the interaction. In the following we reveal the underlying origin for the fragility of the quasiparticle for $k \sim k_c$, showing that it is actually due to the coupling of quasiparticles with the collective plasmon mode.
Figure~\ref{fig:density_plot} presents two-dimensional plots of the spectral function and the imaginary part of the self-energy as a function of momentum and energy for various values of $r_s$. 
The plasmon satellite band indicated by the white arrow arises due to the coupling between electrons and plasmons, which are the collective modes of charge density oscillation. 
Near the Fermi surface (i.e., $\hbar\omega\ll E_\mathrm{F}$ and $k\approx k_\mathrm{F}$), the plasmon satellite band disperses as $\omega(k)\sim \sqrt{k-k_\mathrm{F}}$, and thus is well separated from the quasiparticle band dispersion going linearly in $k-k_\mathrm{F}$. As we move away the Fermi surface, however, the plasmon satellite band disperses toward the renormalized quasiparticle band, eventually merging with it at the critical wave vector $k_c$, where quasiparticles are most fragile. This leads to a coupling between the plasmon collective mode and quasiparticles, and a nonanalytic kink structure appears in the renormalized energy band at the critical wave vector $k_c$, accompanying a rapid increase of the imaginary part of the self-energy as shown in the lower figures of Fig~\ref{fig:density_plot}. Such an increase arises because of the triggering of an additional decay channel via the plasmon emission, which is absent at low energies near the Fermi surface.
This interpretation is consistent with the previous results where $Z_k$ is suppressed to a very small value as it approaches the critical wave vector $k_c$. We also emphasize that the results of Fig.~\ref{fig:density_plot} show why the quasiparticles are most fragile around the critical wave vector $k_c$, and better defined as we move to higher energies away from $k_c$.

\begin{figure*}[!htb]
  \centering
  \includegraphics[width=\linewidth]{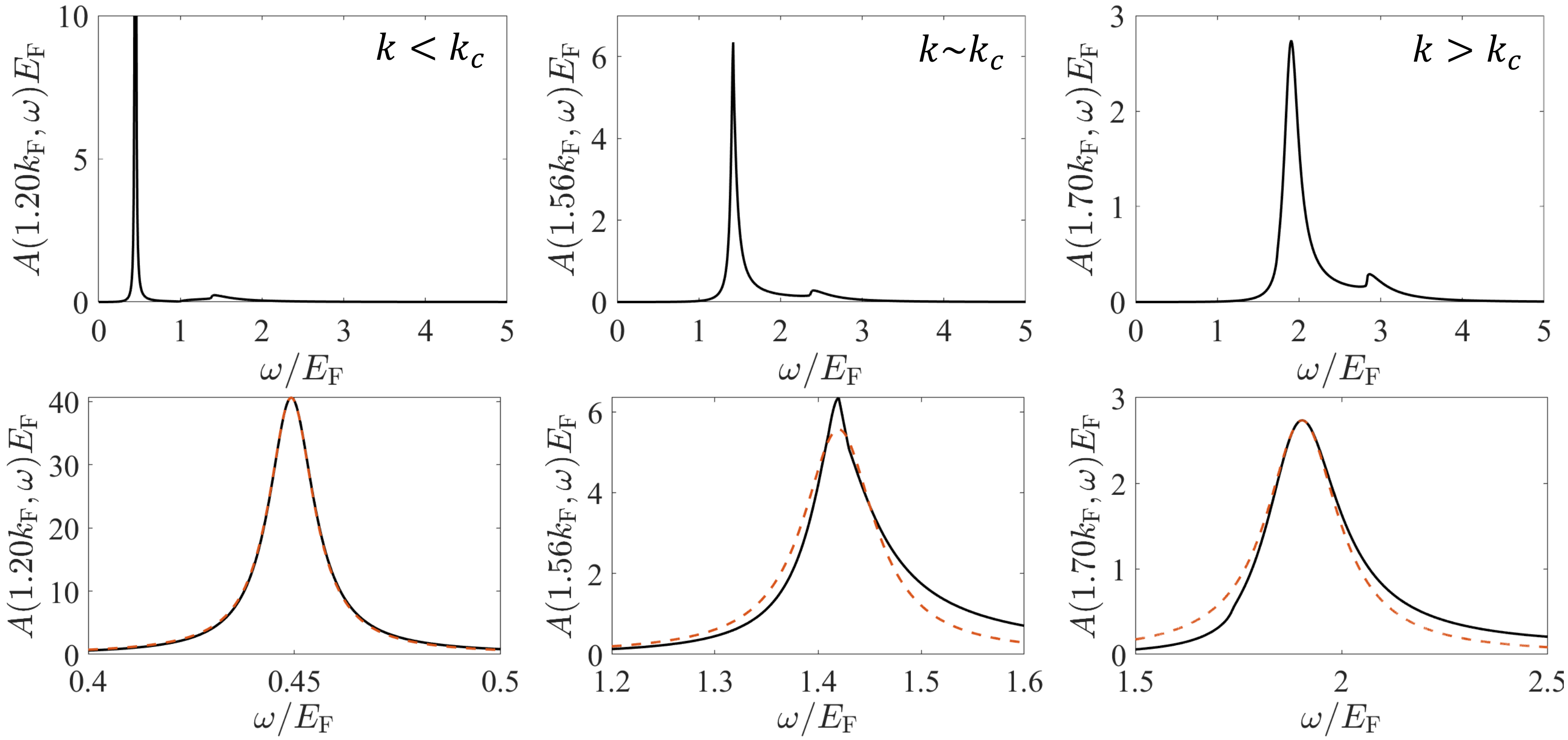}
  \caption{The spectral functions for 3DEG plotted as a function of $\omega$ for three different values of $k$: $k=1.2k_\mathrm{F}<k_c$, $k=1.2k_\mathrm{F}\sim k_c$ and $k=1.2k_\mathrm{F}>k_c$. The figures in the second row present a zoom in of the spectral peak along the best Lorentzian fit indicated by red dashed lines. Here $r_s=1.0$ is used.   }
  \label{fig:3DEG_spectral_functions}
\end{figure*}

In the weakly interacting limit ($r_s\ll1$), the plasmon satellite band simply behaves as a pure plasmon, whose energy dispersion can be obtained by finding the poles of the dielectric function [Eq.~(\ref{eq:diel})], i.e.,
\begin{equation}
    \varepsilon\left[ q,\omega_\mathrm{p}(q)\right]=0,
    \label{eq:diel2}
\end{equation}
where $\omega_\mathrm{p}(q)$ is the plasmon dispersion.
Since the plasmon satellite band starts from the Fermi surface (i.e., $k=k_\mathrm{F}$) as shown in Fig.~\ref{fig:density_plot}, the actual dispersion for the plasmon satellite band should be written as $k_\mathrm{F}$, i.e., $\omega_\mathrm{p}(k-k_\mathrm{F})$.
Similarly, the quasiparticle energy dispersion can be approximated by the noninteracting quadratic energy band dispersion given by $E(k)=\hbar^2k^2/2m$ (see Fig.~\ref{fig:appendix_den_plot_spectral} in Appendix). Thus, by solving the equation $\hbar\omega_\mathrm{p}(q-k_\mathrm{F})=\hbar^2k^2/2m$, we can analytically find the critical wave vector $k_c$ in the weakly interacting limit, which is readily obtained as \cite{Ahn2021}:
\begin{equation}
    \frac{(\widetilde{k}_c-1)^2}{\sqrt{2}r_s} + \frac{(\widetilde{k}_c-1)^3}{4r_s^2}=
    1,
    \label{eq:k_c}
\end{equation}
Here $\widetilde{k}=k/k_\mathrm{F}$. 
Using Eq.~(\ref{eq:k_c}), we find $k_c=1.269k_\mathrm{F}$ and $k_c=1.405k_\mathrm{F}$ for $r_s=0.1$ and $r_s=0.2$, respectively, which are in good agreement with $k_c$ obtained from the direct numerical results in Fig.~\ref{fig:eqp_imag_ratio}. 
Equation.~(\ref{eq:k_c}) is exactly solvable because it is a simple third degree polynomial equation. However, since Eq.~(\ref{eq:k_c}) is correct only for small $r_s$, it is useful to obtain the asymptotic form of $k_c$ in the limit of $r_s\ll1$, which helps us to understand the relation between the critical wave vector $k_c$ and interaction strength $r_s$: 
\begin{equation} \label{eq:asymptotic_k_c}
    \widetilde{k}_c = (2r_s)^{2/3} - \frac{2\sqrt{2}r_s}{3}+ 1 + O(r_s^{4/3}).    
\end{equation} 
This analysis also manifestly clarifies the connection between the fragility of the quasiparticles around $k\sim k_c$ with the threshold triggering of the coupling between 2D plasmons and the electron-hole excitations.


\section{comparison to 3DEG} \label{sec:comparison_to_3DEG}
The validity of the Landau quasiparticle theory significantly depends on dimensionality (For example, there is no Fermi liquid in 1D where any interaction destroys the Fermi surface).
In 3D, the quasiparticle picture is known to work well, and may become fragile only in the very low density limit where the kinetic energy is strongly suppressed by the Coulomb interaction. 
In this section, we investigate the validity of 3D quasiparticle picture.

Figure~\ref{fig:3DEG_spectral_functions} shows the calculated spectral function of 3DEG, and the result of a fit to the Lorentzian line shape right below, using the same leading-order dynamical RPA theory as employed for the 2DEG above. For $k<k_c$, the spectral function manifests the typical quasiparticle behavior, well fitted by a Lorentzian curve. With increasing $k$, we find that there exists a critical wave vector $k_c$, similar to 2DEG, where the shape of the spectral function deviates from the Lorentzian curve. Note, however, that the distortion is nowhere as dramatic as in the 2DEG case, where the spectral peak becomes extremely sharp with an almost arbitrary shape as $k\rightarrow k_c$. 
For $k>k_c$, the spectral function recovers its Lorentzian shape, but with a large width compared to its height, which is again similar to the 2DEG result. 
It is worth noting that the spectral evolution as a function of $k$ from $k<k_c$ to $k>k_c$ is quite similar to that of the 2DEG: the spectral function has a Lorentzian line shape except in the small region around $k=k_c$.   
This shows that the quasiparticle coupling with plasmons occurs also in 3DEG, and thus 3D quasiparticles are somewhat fragile around the critical wave vector $k_c$ where the quasiparticles are coupled with collective plasmon excitations. It is important to note, however, that the coupling is much weaker in 3D than in 2DEG, which can be seen by the small deviation from the Lorentzian line shape, and thus 3D quasiparticles are more stable than 2D quasiparticles. The difference between 2D and 3D situations is, however, quantitative, and not qualitative, with the 3D case showing quantitatively less singular behavior around $k\sim k_c$, but both 2D and 3D manifest fragile quasiparticles around $k\sim k_c$ with the fragility being stronger for 2DEG. Thus, 3D and 2D interacting Fermi liquids are similar qualitatively while differing strongly quantitatively.

\begin{figure}[!htb]
  \centering
  \includegraphics[width=\linewidth]{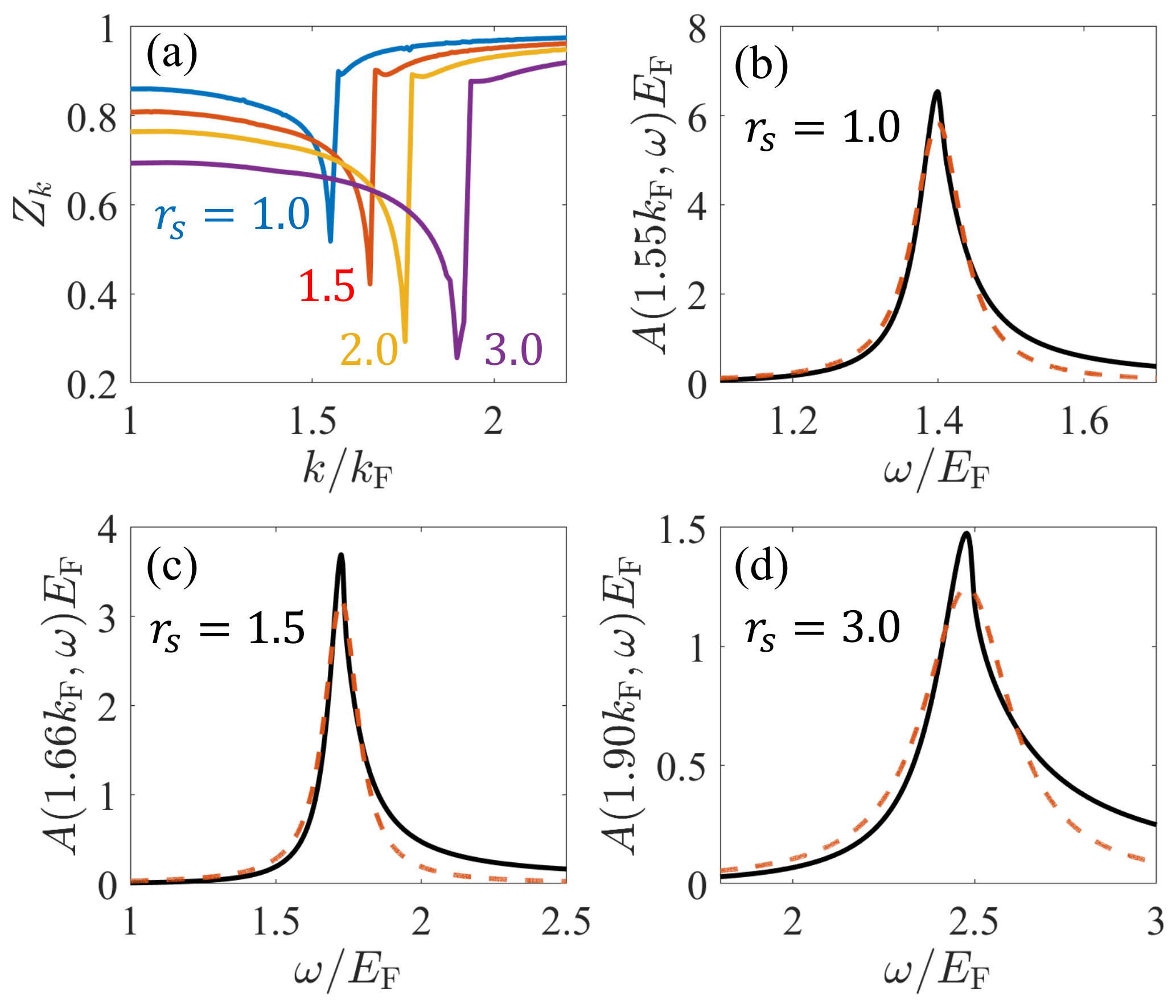}
  \caption{Results for 3D: (a) the renormalization factor $Z_k$ as a function of the momentum $k$ for various values of $r_s=1.0$, $1.5$, $2.0$, and $3.0$. [(b)-(d)] the spectral functions at the critical wave vector $k_c$ plotted as a function of the energy $\omega$ for various values of $r_s=1.0$, $1.5$, and $3.0$. The red dashed line are the best Lorentzian fits to the spectral peak.}
  \label{fig:Zk_plot_3D}
\end{figure}

Figure~\ref{fig:Zk_plot_3D} presents the calculated renormalization factor for 3DEG plotted as a function of the wave vector for various values of $r_s$. For all values of $r_s$, $Z_k$ decreases with increasing $k$, reaching the minimum value at a certain critical wave vector $k_c$ like the 2D $Z_k$ results. Note, however, that $Z_k$ is reduced to a finite value as $k\rightarrow k_c$ instead of vanishing to almost zero as in the 2D results. 
We note that the minimum of the renormalization factor, i.e., $Z_{k_c}$, becomes smaller with increasing $r_s$ unlike the 2D results where the minimum of the renormalization factor is always near zero for any values of $r_s$. In a moderate range of $r_s$, therefore, we expect the quasiparticle picture to work well in 3D at all energies even in the region around the critical wave vector since $Z_k$ does not become particularly small unless $r_s$ is large (where our theory is not accurate).
In Figs~\ref{fig:Zk_plot_3D} (b)-\ref{fig:Zk_plot_3D} (d), we plot the spectral function as a function of energy at the critical wave vector $k=k_c$. Note that with increasing $r_s$, the spectral function deviates strongly from the Lorentzian shape. However, even for a very large $r_s=3.0$, the deviation of the spectral function from the Lorentzian shape is much less dramatic compared to that of 2DEG, showing that 3D quasiparticles are much better defined than 2D quasiparticles although the difference is quantitative.
As we go across $k=k_c$, the renormalization factor exhibits an abrupt jump. Unlike 2DEG, $Z_k<1$ even for $k>k_c$ in 3DEG, which is in contrast to the 2DEG result where the renormalization factor is greater than unity for $k>k_c$. This again shows that the quasiparticle picture is quantitatively more robust in 3DEG than in 2DEG.

Comparing the 2DEG and 3DEG results, we conclude that 2D quasiparticles have both the 1D collective and the 3D quasiparticle aspects: 3D quasiparticles are generally well-defined at all energies, whereas one-dimensional quasiparticles vanish for infinitesimal very small interaction.
For 2DEG, the system is governed by collective excitations around the critical wave vector $k_c$ where the renormalization factor is greatly reduced, while remaining nonzero, and thus, it approaches almost 1D type behavior at $k_c$, but not quite. Thus, 2D is similar to 3D, while quantitatively the quasiparticles are more fragile in 2D than in 3D, but in both systems the quasiparticles exist (except perhaps just at $k_c$ in 2D where $Z_k \sim 0$ at $k=k_c$) unlike in 1D where no quasiparticles survive at any momentum, including $k=k_F$, for any interaction.

\begin{figure}[!htb]
  \centering
  \includegraphics[width=\linewidth]{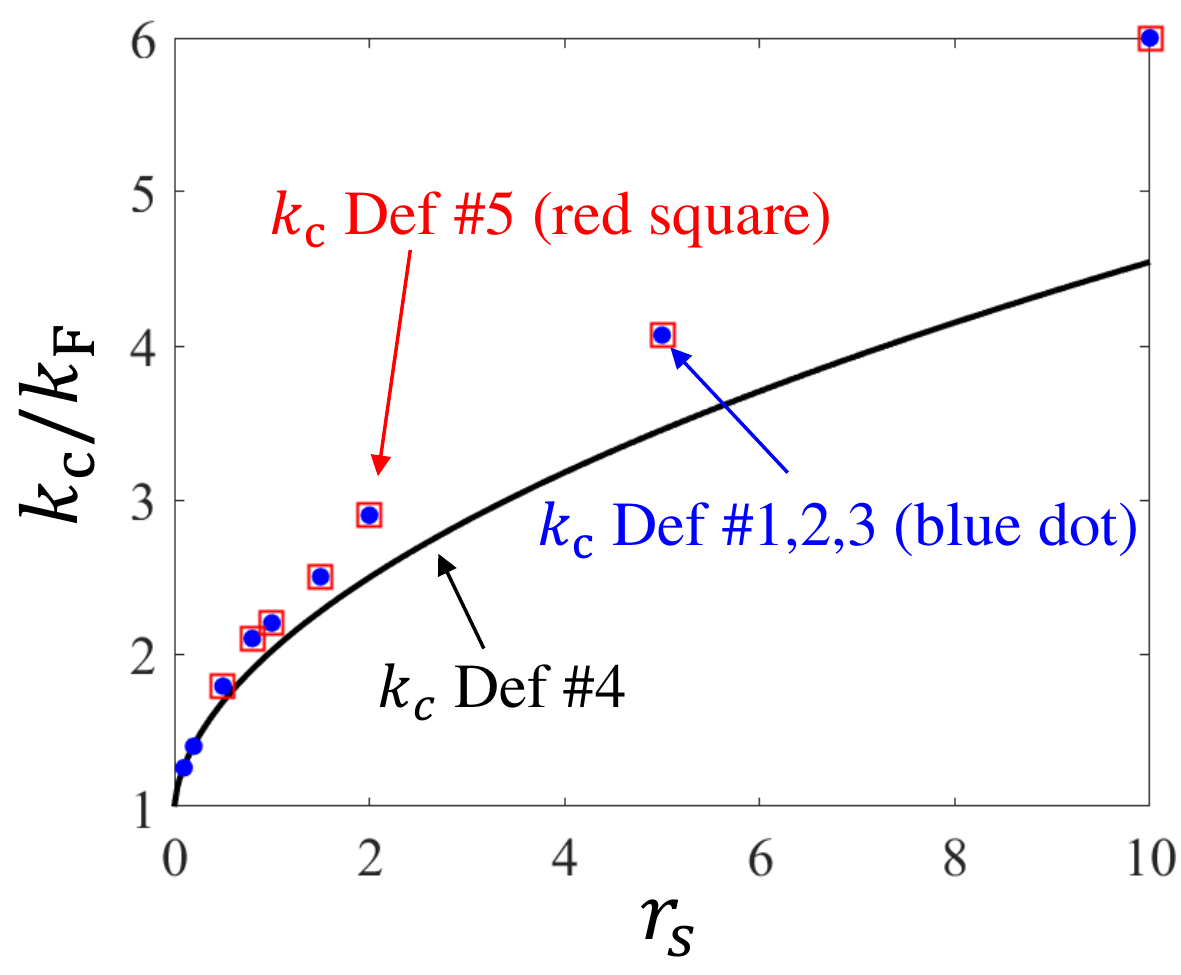}
  \caption{Plot of the critical 2D wave vector $k_c$ obtained using different independent definitions listed in the main text. }
  \label{fig:kc_plot}
\end{figure}

\section{Discussion and Conclusion} \label{sec:discussion_and_conclusion}
We have theoretically studied the zero-temperature quasiparticle properties of an interacting 2DEG using the leading-order expansion in the RPA-screened dynamical Coulomb interaction, a theory which is exact in the high-density $r_s\ll 1$ limit, and have compared our results to the corresponding 3DEG case, also commenting on how the 2D situation differs from the 1D Luttinger liquid system. We have obtained the real and imaginary parts of the dynamical self-energy, the quasiparticle energies in the on-shell and off-shell approximations, the energy and momentum-dependent quasiparticle spectral function, the quasiparticle renormalization factor, and the effective mass for various values of $r_s$, commenting on both the weakly interacting $r_s<1$ and the strongly interacting $r_s>1$ regimes. The qualitative results are the same for all values of $r_s$, giving confidence that, although our results are strictly valid only for the $r_s\ll 1$ regime, our qualitative conclusions are valid for all $r_s$ unless there is an $r_s$-driven quantum phase transition. 
In fact, 3D metals have $r_s\sim 4$-$6$, and it is well-known that the RPA many-body theory used in the current work provides a reasonable description at metallic densities \cite{Rice1965, abrikosov2012methods}. 
The focus has been investigating the extent to which the interacting 2DEG is a Fermi liquid with well-defined stable quasiparticles. We find that not only the 2DEG has well-defined quasiparticles at all values of $r_s$, these quasiparticles are stable at all values of momentum and energy, even very far above the Fermi surface, except perhaps a critical momentum $k_c$ where the quasiparticles are extremely fragile, and may actually be suppressed. We also find that at very high energies, the 2D quasiparticles become essentially like noninteracting free particles (i.e., renormalization factor $\sim 1$), but this approach to the noninteracting limit curiously happens in a rather intriguing manner with the renormalization factor being slightly above unity at momentum well above $k_c$.

We emphasize that the critical wave vector $k_c$, where the quasiparticles are most fragile, has been obtained throughout the paper using the following different definitions which are completely independent of each other: $k_c$ is defined as a wave vector where the followings are true.
\begin{enumerate}
    \item $E_\mathrm{QP}/|\mathrm{Im}\Sigma|$ is the minimum both in off-shell and on-shell approximations.
    \item The renormalization factor $Z_k$ vanishes.
    \item The spectral function is most non-Lorentzian with multiple solutions of Dyson's equation.
    \item The plasmon dispersion just crosses into the single-particle dispersion enabling strong quasiparticle decay through plasmon emission so the imaginary part of the self-energy has a sudden increase.
\end{enumerate}

Each of these definitions is closely and equivalently related to the validity condition for the quasiparticle picture.
Definition 1 is from the most fundamental condition for the quasiparticle concept to be valid according to the Landau Fermi liquid theory as discussed in Sec.~\ref{sec:quasiparticle}.
The physical meaning of definition 2 is that at the critical wave vector $k_c$, most of the spectral weight is incoherent so the coherent quasiparticle weight is entirely suppressed leading to a breakdown of the quasiparticle picture, but just at a set of measure zero, i.e., only at $k=k_c$. 
Definition 3 implies that the interacting spectral function cannot be approximated by a coherent single-particle spectral function, which is Lorentzian in the form of $A\sim \eta/(E_\mathrm{QP}^2+\eta^2)$, and thus a strong deviation from the Lorentzian shape implies that the interacting excitations cannot be described as quasiparticles.
Remarkably, the alternate definitions 1, 2 and 3 give exactly the same critical wave vector in our theory as shown in Fig.~\ref{fig:kc_plot}. This consistency shows that 2D quasiparticles are fragile in a small regime around $k_c$ or more likely, just at $k=k_c$, the momentum threshold for plasmon emission by quasiparticles.
Definition 4 reveals why the Fermi liquid description becomes inapplicable at the critical wave vector $k_c$. The black line in Fig~\ref{fig:kc_plot} represents the critical wave vector given by definition 4, which is in good agreement with those from the three definitions 1-3 in the weakly interacting limit (up to $r_s\sim 0.2$) where our theory is essentially exact.

Other than the criteria discussed above, it is quite common in the literature to use only the broadening of the spectral peak to ascertain the validity of the quasiparticle picture. According to this criterion, if the broadening is large enough, the quasiparticle picture breaks down. In this context, we can define, using a fifth independent criterion, the critical wave vector $k_c$ as
\begin{enumerate}
\setcounter{enumi}{4}
     \item the smallest wave vector where where the spectral width is larger than the spectral height in the natural unit of $E_\mathrm{F}$, i.e., $\frac{AE_\mathrm{F}}{\Delta \hbar\omega / E_\mathrm{F}}>1$.
 \end{enumerate}
In Fig.~\ref{fig:kc_plot} the critical wave vectors $k_c$ obtained using height versus width criterion 5 (red squares) for various values of $r_s$ are plotted as red squares. For small $r_s$, since the broadening of the spectral function is very small at all energy scales, as shown in Fig.~\ref{fig:spectral_function_at_large_k}, there exists no critical wave vector $k_c$ (at low $r_s$) following definition 5. 
For $r_s$ larger than $\sim 0.8$, there exists a $k_c$ where the spectral peak becomes broad enough so that criterion 5 is satisfied. 
Although criterion 5 has a different context (and is not obviously related to the first four criteria for quasiparticle stability used above), it is noteworthy that for larger $r_s$ ($>0.8$), it gives essentially the same critical $k_c$ as the first four criteria do.
We note that criterion 5 is somewhat arbitrary since the spectral height and width have fundamentally different (in fact, inverse of each other) units (1/energy and energy respectively), and cannot therefore be compared directly as, for example, the quasiparticle energy and the broadening in criterion 1 can be. Although dimensionalizing them using $E_\mathrm{F}$ may sound reasonable and it is comforting to see that it provides a $k_c$ which is consistent with $k_c$ values obtained from the first four criteria, we must keep in mind that this fifth criterion is quite arbitrary since the height and width of the spectral function are fundamentally not comparable even if such a comparison seems intuitive in defining stable quasiparticles.
We emphasize, however, that this operational definition is enough to conclude that definition 5 gives the identical critical wave vectors $k_c$ as the other definitions 1-4.

We note that the 3DEG results are somewhat similar to the 2D results except that the quasiparticles are not completely suppressed at a critical momentum $k_c$ in 3D as they are in the 2DEG although a critical wave vector $k_c$ exists in both cases with the $k_c$ in 3D manifesting a less stable quasiparticle with the renormalization factor showing a minima without going to zero. The difference between 2D and 3D is thus quantitative, and we conclude that they are similar except that the quasiparticle appears to be completely suppressed at $k=k_c$ in 2D in contrast to 3D where it is just strongly suppressed. By contrast, the 1D interacting system formally has a vanishing quasiparticle renormalization factor at all momenta, including even the Fermi momentum, indicating a nonexistence of a 1D Fermi liquid. In some sense, we can say that in 1D all momenta become critical with all $k$ values being $k_c$ (including $k_\mathrm{F}$) whereas, in 2D, $k_c$ is a set of measure zero, being precisely one momentum value above $k_\mathrm{F}$. In 3D, there is a well-defined $k_c$ also except here $k_c$ only implies a momentum where the quasiparticle weight is a minimum without vanishing, so the quasiparticle is always stable. Thus, in 1D the quasiparticles do not exist at any momentum (including the Fermi momentum) whereas in 3D quasiparticles are stable at all momenta whereas in 2D, the quasiparticles are stable at all momenta except possibly at one critical momentum. Although one does not usually think of 1D interacting fermions in terms of perturbative arguments since there is no small parameter in 1D, it can be shown that for Coulomb interacting systems, the same diagrams as the ones calculated in the current work (i.e., the infinite ring diagrams in an expansion in the leading-order dynamical RPA theory) immediately lead to the conclusion that there are no quasiparticles in 1D and there is no interacting Fermi surface \cite{Hu1993Many}.

We must also emphasize very clearly that although we have mostly discussed the intriguing role of the critical momentum $k_c$ for the interacting 2DEG, mainly because of the unanticipated total suppression of the 2D quasiparticle spectral weight at $k_c$, our main finding is that 2D quasiparticles are stable and well-defined at all momenta and energies for all $r_s$ values (except of course at $k=k_c$), and although the incoherent contribution to the interacting spectral function is often appreciable, there is always a stable coherent piece well-described by a Lorentzian, with $E_\mathrm{QP}$ being larger than the imaginary part of the self-energy at all momenta except for $k\sim k_c$. We also recover the quasiparticle renormalization factor approaching unity at large energy since interaction effects should disappear at very high momentum and energy. Quasiparticles are perturbatively stable in 2D (just as they are in 3D) in the presence of Coulomb interactions, except for perhaps one precise momentum $k_c$. Our work provides the quantitative details for the renormalization effects in 2DEG and 3DEG arising from Coulomb interaction appropriate for simple metals and doped 2D semiconductors.

One may wonder what our $T=0$ theory implies for the finite temperature situation with respect to the validity or not of the finite-temperature Fermi liquid theory and the quasiparticle picture. Very crudely speaking we could reinterpret the electron energy away from the Fermi surface as a temperature in the corresponding finite temperature theory, and conclude qualitatively that the Fermi liquid theory and the quasiparticle picture remain applicable for very large temperatures above the Fermi temperature, even $T>T_\mathrm{F}$, since our $T=0$ theory indicates the validity of the Fermi liquid picture for energies $E>E_\mathrm{F}$ away from the Fermi surface. This indeed seems to be the situation as recent works have explored the validity of the Fermi liquid quasiparticle picture at finite temperatures, finding that the imaginary part of the self-energy remains smaller than the quasiparticle energies up to arbitrarily high temperatures, $T\gg T_\mathrm{F}$ \cite{Sarma2021know}. One should be careful, however, because in some sense there are no well-defined quasiparticles at finite temperatures even for a noninteracting electron gas as the Fermi distribution function is continuous at any finite temperature, and the Fermi surface discontinuity disappears trivially at all finite temperatures. We refer to Ref.~\cite{Sarma2021know} for more details on the finite temperature RPA self-energy calculations.

To conclude, we have revisited the quasiparticle Landau theory, providing a comprehensive treatment of 2D quasiparticles for Coulomb interactions from low energies near the Fermi surface up to high energies where the quasiparticle behaves essentially as a free particle. We investigate the domain of validity of the quasiparticle Landau theory by explicitly calculating the real and imaginary parts of the self-energies, the spectral functions, and the renormalization factor within the leading-order dynamical theory which is exact at high densities, which has been highly successful in obtaining reliable quasiparticle properties of interacting systems. 
In contrary to the popular belief that the quasiparticle picture becomes ill-defined far away from the Fermi surface (i.e., quasiparticles exist only at low energies near the Fermi surface), we find that quasiparticles are robust up to high energies, showing that the quasiparticle energy is always larger than the imaginary part of the self-energy essentially everywhere (except for a set of measure zero at a critical momentum $k_c$) and for all Coulomb coupling strengths. We find, however, that there exists a small region around a critical wave vector $k_c$ where the quasiparticle picture becomes fragile (and probably fails completely) due to the strong coupling of quasiparticles with the plasmon collective modes. This also happens in 3DEG, but much less prominently than in 2DEG. 
Our conclusion is therefore that although Coulomb interaction effects are strong in the 2DEG (much stronger than in 3DEG), the interacting system is qualitatively similar to 3D metals where the quasiparticle picture and the Landau Fermi liquid theory apply at all energies and momenta, not just close to the Fermi surface. This provides a direct explanation for why band theories are so extremely successful in both 2D and 3D unless some nonperturbative strong correlation effects (e.g. Mott transition in a lattice or Wigner transition in the continuum) intervene causing a quantum phase transition invalidating our perturbative theoretic results.

Finally, we comment on possible improvements of our theory, which is exact only for the weak-coupling $r_s\ll 1$ high-density limit. One obvious question in this context is if it is feasible to go to the next-to-the-leading-order in the dynamically screened Coulomb interaction since the current theory is the leading-order theory. Our answer is that it is essentially impossible to carry out any systematic higher-order diagrammatic many-body field theory calculations along the line of the systematic perturbative expansion in $r_s$ we carried out here. The reason is that the next-to-the-leading-order perturbation theory involves far too many diagrams because one must take functional derivatives of the electron-hole bubbles along with the obvious higher-order terms in the RPA-screened interactions themselves. These terms in general are multidimensional singular integrals which cannot be calculated except by making drastic simplifying (and uncontrolled) approximations, e.g., $1/N$ approximation \cite{Hofmann2014why}, which would not apply to any physical 2DEG. Such approximations must be entirely numerical using demanding Monte Carlo techniques, and detailed results for the spectral function similar to what is presented in the current work are then impossible. One may resort entirely to nonperturbative numerical quantum Monte Carlo techniques which are often successful in obtaining ground state properties, but are typically not useful in answering fundamental issues of principle (e.g. the detailed structure of the interacting spectral function) discussed in our work.

\section{Acknowledgement} \label{sec:acknowledgement}
This work is supported by the Laboratory for Physical Sciences.

\appendix* \label{sec:appendix}
\section{ Additional results for different $r_s$ values }
\begin{figure*}
     \centering
     \subfloat[][The upper five figures show the calculated real and imaginary part of self-energies near the Fermi surface ($k=1.1k_\mathrm{F}$), around the critical wave vector ($k = 1.2k_\mathrm{F}$, $1.3k_\mathrm{F}$, and $1.4k_\mathrm{F}$) and at high energies far away from the Fermi surface ($k = 1.6k_\mathrm{F}$). For visual clarity, $|\mathrm{Im}\Sigma|$ is plotted instead of $\mathrm{Im}\Sigma$. The straight dashed lines are given by $\hbar\omega - \varepsilon_{\bm k} + E_\mathrm{F}$, whose intersection with $\mathrm{Re}\Sigma$ corresponds to the solutions of the Dyson's equation giving sharp peaks of the spectral functions plotted below. The figures in the third row show the zoom-in of the spectral peak along with the best fit curve by a Lorentzian distribution (dahsed red). Here we set $r_s=0.2$, and $\omega$ is measured from the interacting Fermi energy.]{\includegraphics[width=\linewidth]{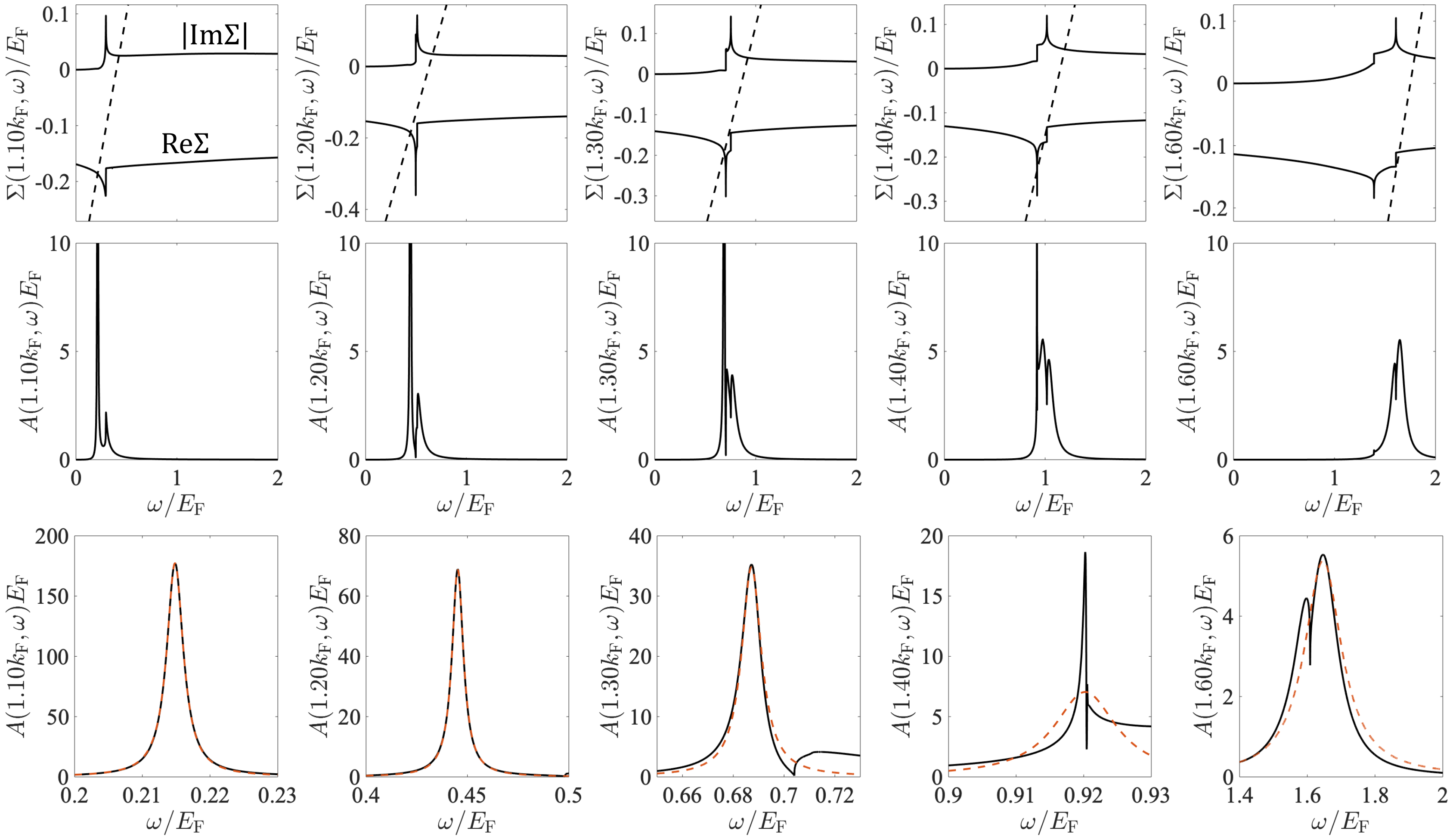}\label{fig:spectral_functions_0.2}}
     
     \subfloat[][Same as Fig (a) but for $r_s=2.0$.]{\includegraphics[width=\linewidth]{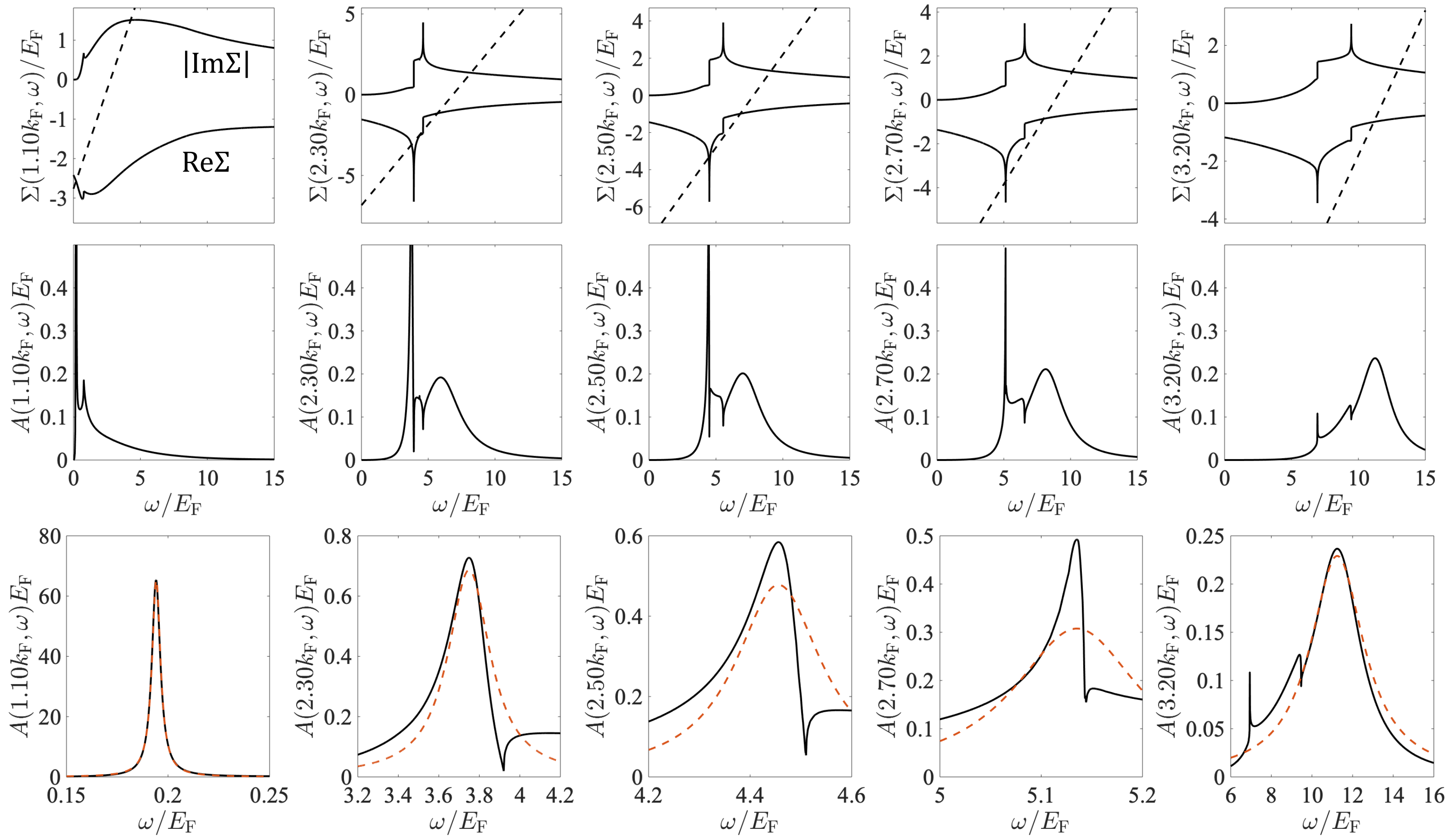}\label{fig:spectral_functions_2.0}}
    \caption{Self-energies and spectral functions for (a) small and (b) large values of $r_s$ along with the best Lorentzian fit to the spectral peak.}
    \label{fig:spectral_functions_Appendix}
\end{figure*}

\begin{figure*}[!htb]
  \centering
  \includegraphics[width=\linewidth]{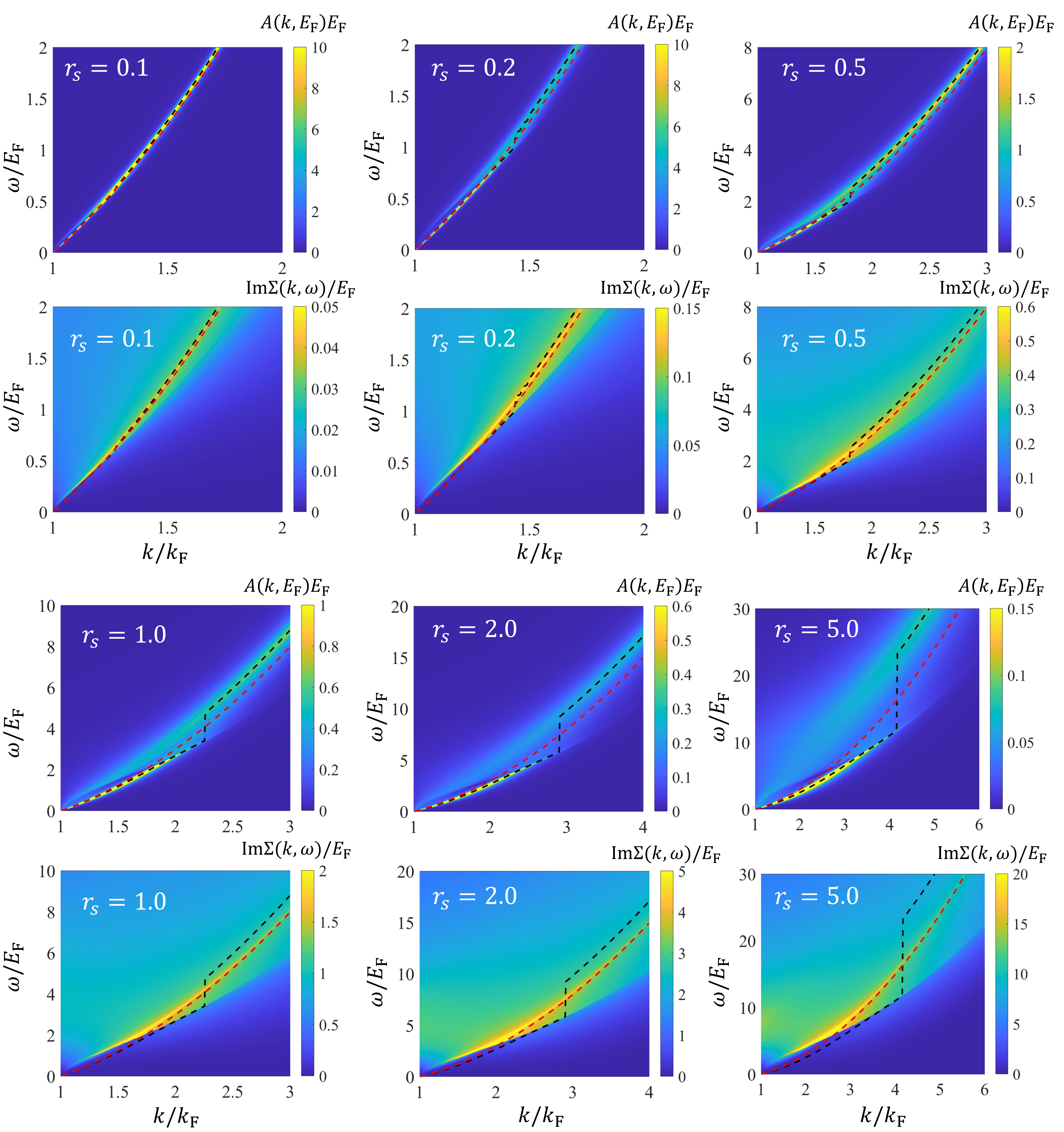}
  \caption{Two-dimensional false-color plot of the spectral function and the imaginary part of the self-energy for small $r_s=0.1$, $0.2$, $0.5$ (upper two rows), and for large $r_s=1.0$, $2.0$, $5.0$ (lower two rows). The dashed black and red lines represent the renormalized and bare energy band dispersion, respectively, which are in good agreement for small $r_s$. }
  \label{fig:appendix_den_plot_spectral}
\end{figure*}

In this appendix, we present the results of the main text for different values of $r_s$ to show that all the key qualitative features discussed in the main text remain qualitatively the same regardless of the interaction strength.
Figure~\ref{fig:spectral_functions_Appendix} shows results corresponding to the Fig.~\ref{fig:spectral_functions} results of the main text for smaller $r_s=0.2$ and larger $r_s=2.0$. It is worth noting that even for small $r_s=0.2$ where our leading-order approximation is known to be exact, the spectral function develops a sharp non-Lorentzian peak approaching the critical wave vector $k_c$. 
Figure~\ref{fig:spectral_Aqc_Ainc_Appendix} shows the evolution of the spectral function with increasing momentum $k$ for various values of $r_s$. The figure shows that near the critical wave vector the quasiparticle peak is split into the coherent and incoherent parts regardless of the value of $r_s$.  
We also present in Fig.~\ref{fig:appendix_den_plot_spectral} the two-dimensional false-color plot of the spectral functions and the imaginary part of the self-energy along with the renormalized (black-dashed) and bare (red-dashed) energy dispersions. It is important to note that for all values of $r_s$ ranging from $0.1$ to $5.0$ the collective plasmon mode crosses into the single-particle renormalized energy dispersion at the critical wave vector $k_c$, showing that the coupling of quasiparticles with plasmon collective modes occurs regardless of the interaction strength. Also note that the renormalized energy dispersion can be well approximated by the parabolic energy dispersion, which numerically justifies the derivation of Eq.~(\ref{eq:asymptotic_k_c}).
All the results in this appendix show that the discussions and conclusions in the main text are qualitatively valid in both weakly and strongly interacting regimes.

\bibliographystyle{apsrev4-2}
\bibliography{ref}

\begin{thebibliography}{33}%
\makeatletter
\providecommand \@ifxundefined [1]{%
 \@ifx{#1\undefined}
}%
\providecommand \@ifnum [1]{%
 \ifnum #1\expandafter \@firstoftwo
 \else \expandafter \@secondoftwo
 \fi
}%
\providecommand \@ifx [1]{%
 \ifx #1\expandafter \@firstoftwo
 \else \expandafter \@secondoftwo
 \fi
}%
\providecommand \natexlab [1]{#1}%
\providecommand \enquote  [1]{``#1''}%
\providecommand \bibnamefont  [1]{#1}%
\providecommand \bibfnamefont [1]{#1}%
\providecommand \citenamefont [1]{#1}%
\providecommand \href@noop [0]{\@secondoftwo}%
\providecommand \href [0]{\begingroup \@sanitize@url \@href}%
\providecommand \@href[1]{\@@startlink{#1}\@@href}%
\providecommand \@@href[1]{\endgroup#1\@@endlink}%
\providecommand \@sanitize@url [0]{\catcode `\\12\catcode `\$12\catcode
  `\&12\catcode `\#12\catcode `\^12\catcode `\_12\catcode `\%12\relax}%
\providecommand \@@startlink[1]{}%
\providecommand \@@endlink[0]{}%
\providecommand \url  [0]{\begingroup\@sanitize@url \@url }%
\providecommand \@url [1]{\endgroup\@href {#1}{\urlprefix }}%
\providecommand \urlprefix  [0]{URL }%
\providecommand \Eprint [0]{\href }%
\providecommand \doibase [0]{https://doi.org/}%
\providecommand \selectlanguage [0]{\@gobble}%
\providecommand \bibinfo  [0]{\@secondoftwo}%
\providecommand \bibfield  [0]{\@secondoftwo}%
\providecommand \translation [1]{[#1]}%
\providecommand \BibitemOpen [0]{}%
\providecommand \bibitemStop [0]{}%
\providecommand \bibitemNoStop [0]{.\EOS\space}%
\providecommand \EOS [0]{\spacefactor3000\relax}%
\providecommand \BibitemShut  [1]{\csname bibitem#1\endcsname}%
\let\auto@bib@innerbib\@empty
\bibitem [{\citenamefont {Landau}(1957)}]{Landau1957}%
  \BibitemOpen
  \bibfield  {author} {\bibinfo {author} {\bibfnamefont {L.}~\bibnamefont
  {Landau}},\ }\href@noop {} {\bibfield  {journal} {\bibinfo  {journal} {Sov.
  phys. JETP}\ }\textbf {\bibinfo {volume} {5}},\ \bibinfo {pages} {101}
  (\bibinfo {year} {1957})}\BibitemShut {NoStop}%
\bibitem [{\citenamefont {Landau}(1959)}]{Landau1959}%
  \BibitemOpen
  \bibfield  {author} {\bibinfo {author} {\bibfnamefont {L.}~\bibnamefont
  {Landau}},\ }\href@noop {} {\bibfield  {journal} {\bibinfo  {journal} {Sov.
  Phys. JETP}\ }\textbf {\bibinfo {volume} {8}},\ \bibinfo {pages} {70}
  (\bibinfo {year} {1959})}\BibitemShut {NoStop}%
\bibitem [{\citenamefont {Baym}\ and\ \citenamefont
  {Pethick}(2008)}]{Baym2008Landau}%
  \BibitemOpen
  \bibfield  {author} {\bibinfo {author} {\bibfnamefont {G.}~\bibnamefont
  {Baym}}\ and\ \bibinfo {author} {\bibfnamefont {C.}~\bibnamefont {Pethick}},\
  }\href@noop {} {\emph {\bibinfo {title} {Landau Fermi-liquid theory: concepts
  and applications}}}\ (\bibinfo  {publisher} {John Wiley \& Sons},\ \bibinfo
  {year} {2008})\BibitemShut {NoStop}%
\bibitem [{\citenamefont {Pines}(2018)}]{Pines2018Theory}%
  \BibitemOpen
  \bibfield  {author} {\bibinfo {author} {\bibfnamefont {D.}~\bibnamefont
  {Pines}},\ }\href@noop {} {\emph {\bibinfo {title} {Theory of Quantum
  Liquids: Normal Fermi Liquids}}}\ (\bibinfo  {publisher} {CRC Press},\
  \bibinfo {year} {2018})\BibitemShut {NoStop}%
\bibitem [{\citenamefont {Quinn}\ and\ \citenamefont
  {Ferrell}(1958)}]{quinn1958electron}%
  \BibitemOpen
  \bibfield  {author} {\bibinfo {author} {\bibfnamefont {J.~J.}\ \bibnamefont
  {Quinn}}\ and\ \bibinfo {author} {\bibfnamefont {R.~A.}\ \bibnamefont
  {Ferrell}},\ }\href@noop {} {\bibfield  {journal} {\bibinfo  {journal}
  {Physical Review}\ }\textbf {\bibinfo {volume} {112}},\ \bibinfo {pages}
  {812} (\bibinfo {year} {1958})}\BibitemShut {NoStop}%
\bibitem [{\citenamefont {Rice}(1965)}]{Rice1965}%
  \BibitemOpen
  \bibfield  {author} {\bibinfo {author} {\bibfnamefont {T.~M.}\ \bibnamefont
  {Rice}},\ }\href {https://doi.org/10.1016/0003-4916(65)90234-4} {\bibfield
  {journal} {\bibinfo  {journal} {Annals of Physics}\ }\textbf {\bibinfo
  {volume} {31}},\ \bibinfo {pages} {100} (\bibinfo {year} {1965})}\BibitemShut
  {NoStop}%
\bibitem [{\citenamefont {Abrikosov}\ \emph {et~al.}(2012)\citenamefont
  {Abrikosov}, \citenamefont {Gorkov},\ and\ \citenamefont
  {Dzyaloshinski}}]{abrikosov2012methods}%
  \BibitemOpen
  \bibfield  {author} {\bibinfo {author} {\bibfnamefont {A.~A.}\ \bibnamefont
  {Abrikosov}}, \bibinfo {author} {\bibfnamefont {L.~P.}\ \bibnamefont
  {Gorkov}},\ and\ \bibinfo {author} {\bibfnamefont {I.~E.}\ \bibnamefont
  {Dzyaloshinski}},\ }\href@noop {} {\emph {\bibinfo {title} {Methods of
  quantum field theory in statistical physics}}}\ (\bibinfo  {publisher}
  {Courier Corporation},\ \bibinfo {year} {2012})\BibitemShut {NoStop}%
\bibitem [{\citenamefont {Jalabert}\ and\ \citenamefont
  {Das~Sarma}(1989)}]{Jalabert1989}%
  \BibitemOpen
  \bibfield  {author} {\bibinfo {author} {\bibfnamefont {R.}~\bibnamefont
  {Jalabert}}\ and\ \bibinfo {author} {\bibfnamefont {S.}~\bibnamefont
  {Das~Sarma}},\ }\href {https://doi.org/10.1103/physrevb.40.9723} {\bibfield
  {journal} {\bibinfo  {journal} {Physical Review B}\ }\textbf {\bibinfo
  {volume} {40}},\ \bibinfo {pages} {9723} (\bibinfo {year}
  {1989})}\BibitemShut {NoStop}%
\bibitem [{\citenamefont {Fetter}\ and\ \citenamefont
  {Walecka}(2012)}]{fetter2012quantum}%
  \BibitemOpen
  \bibfield  {author} {\bibinfo {author} {\bibfnamefont {A.~L.}\ \bibnamefont
  {Fetter}}\ and\ \bibinfo {author} {\bibfnamefont {J.~D.}\ \bibnamefont
  {Walecka}},\ }\href@noop {} {\emph {\bibinfo {title} {Quantum theory of
  many-particle systems}}}\ (\bibinfo  {publisher} {Courier Corporation},\
  \bibinfo {year} {2012})\BibitemShut {NoStop}%
\bibitem [{\citenamefont {Feldman}\ \emph
  {et~al.}(2004{\natexlab{a}})\citenamefont {Feldman}, \citenamefont
  {Kn{\"o}rrer},\ and\ \citenamefont {Trubowitz}}]{feldman2004twoPart1}%
  \BibitemOpen
  \bibfield  {author} {\bibinfo {author} {\bibfnamefont {J.}~\bibnamefont
  {Feldman}}, \bibinfo {author} {\bibfnamefont {H.}~\bibnamefont
  {Kn{\"o}rrer}},\ and\ \bibinfo {author} {\bibfnamefont {E.}~\bibnamefont
  {Trubowitz}},\ }\href@noop {} {\bibfield  {journal} {\bibinfo  {journal}
  {Communications in mathematical physics}\ }\textbf {\bibinfo {volume}
  {247}},\ \bibinfo {pages} {1} (\bibinfo {year}
  {2004}{\natexlab{a}})}\BibitemShut {NoStop}%
\bibitem [{\citenamefont {Feldman}\ \emph
  {et~al.}(2004{\natexlab{b}})\citenamefont {Feldman}, \citenamefont
  {Kn{\"o}rrer},\ and\ \citenamefont {Trubowitz}}]{feldman2004twoPart2}%
  \BibitemOpen
  \bibfield  {author} {\bibinfo {author} {\bibfnamefont {J.}~\bibnamefont
  {Feldman}}, \bibinfo {author} {\bibfnamefont {H.}~\bibnamefont
  {Kn{\"o}rrer}},\ and\ \bibinfo {author} {\bibfnamefont {E.}~\bibnamefont
  {Trubowitz}},\ }\href@noop {} {\bibfield  {journal} {\bibinfo  {journal}
  {Communications in mathematical physics}\ }\textbf {\bibinfo {volume}
  {247}},\ \bibinfo {pages} {49} (\bibinfo {year}
  {2004}{\natexlab{b}})}\BibitemShut {NoStop}%
\bibitem [{\citenamefont {Feldman}\ \emph
  {et~al.}(2004{\natexlab{c}})\citenamefont {Feldman}, \citenamefont
  {Kn{\"o}rrer},\ and\ \citenamefont {Trubowitz}}]{feldman2004twoPart3}%
  \BibitemOpen
  \bibfield  {author} {\bibinfo {author} {\bibfnamefont {J.}~\bibnamefont
  {Feldman}}, \bibinfo {author} {\bibfnamefont {H.}~\bibnamefont
  {Kn{\"o}rrer}},\ and\ \bibinfo {author} {\bibfnamefont {E.}~\bibnamefont
  {Trubowitz}},\ }\href@noop {} {\bibfield  {journal} {\bibinfo  {journal}
  {Communications in mathematical physics}\ }\textbf {\bibinfo {volume}
  {247}},\ \bibinfo {pages} {113} (\bibinfo {year}
  {2004}{\natexlab{c}})}\BibitemShut {NoStop}%
\bibitem [{\citenamefont {Tomonaga}(1950)}]{Tomonaga1950}%
  \BibitemOpen
  \bibfield  {author} {\bibinfo {author} {\bibfnamefont {S.-i.}\ \bibnamefont
  {Tomonaga}},\ }\href {https://doi.org/10.1143/ptp/5.4.544} {\bibfield
  {journal} {\bibinfo  {journal} {Progress of Theoretical Physics}\ }\textbf
  {\bibinfo {volume} {5}},\ \bibinfo {pages} {544} (\bibinfo {year}
  {1950})}\BibitemShut {NoStop}%
\bibitem [{\citenamefont {Luttinger}(1963)}]{Luttinger1963}%
  \BibitemOpen
  \bibfield  {author} {\bibinfo {author} {\bibfnamefont {J.~M.}\ \bibnamefont
  {Luttinger}},\ }\href {https://doi.org/10.1063/1.1704046} {\bibfield
  {journal} {\bibinfo  {journal} {Journal of Mathematical Physics}\ }\textbf
  {\bibinfo {volume} {4}},\ \bibinfo {pages} {1154} (\bibinfo {year} {1963})},\
  \Eprint {https://arxiv.org/abs/https://doi.org/10.1063/1.1704046}
  {https://doi.org/10.1063/1.1704046} \BibitemShut {NoStop}%
\bibitem [{\citenamefont {Zheng}\ and\ \citenamefont
  {Das~Sarma}(1996)}]{Zheng1996}%
  \BibitemOpen
  \bibfield  {author} {\bibinfo {author} {\bibfnamefont {L.}~\bibnamefont
  {Zheng}}\ and\ \bibinfo {author} {\bibfnamefont {S.}~\bibnamefont
  {Das~Sarma}},\ }\href {https://doi.org/10.1103/PhysRevB.53.9964} {\bibfield
  {journal} {\bibinfo  {journal} {Phys. Rev. B}\ }\textbf {\bibinfo {volume}
  {53}},\ \bibinfo {pages} {9964} (\bibinfo {year} {1996})}\BibitemShut
  {NoStop}%
\bibitem [{\citenamefont {{Polchinski}}(1992)}]{Polchinski1992Effective}%
  \BibitemOpen
  \bibfield  {author} {\bibinfo {author} {\bibfnamefont {J.}~\bibnamefont
  {{Polchinski}}},\ }\href@noop {} {\bibfield  {journal} {\bibinfo  {journal}
  {arXiv e-prints}\ ,\ \bibinfo {eid} {hep-th/9210046}} (\bibinfo {year}
  {1992})},\ \Eprint {https://arxiv.org/abs/hep-th/9210046}
  {arXiv:hep-th/9210046 [hep-th]} \BibitemShut {NoStop}%
\bibitem [{\citenamefont {Shankar}(1994)}]{Shankar1994Renormalization}%
  \BibitemOpen
  \bibfield  {author} {\bibinfo {author} {\bibfnamefont {R.}~\bibnamefont
  {Shankar}},\ }\href {https://doi.org/10.1103/RevModPhys.66.129} {\bibfield
  {journal} {\bibinfo  {journal} {Rev. Mod. Phys.}\ }\textbf {\bibinfo {volume}
  {66}},\ \bibinfo {pages} {129} (\bibinfo {year} {1994})}\BibitemShut
  {NoStop}%
\bibitem [{\citenamefont {Gell-Mann}\ and\ \citenamefont
  {Brueckner}(1957)}]{GellMann1957}%
  \BibitemOpen
  \bibfield  {author} {\bibinfo {author} {\bibfnamefont {M.}~\bibnamefont
  {Gell-Mann}}\ and\ \bibinfo {author} {\bibfnamefont {K.~A.}\ \bibnamefont
  {Brueckner}},\ }\href {https://doi.org/10.1103/PhysRev.106.364} {\bibfield
  {journal} {\bibinfo  {journal} {Phys. Rev.}\ }\textbf {\bibinfo {volume}
  {106}},\ \bibinfo {pages} {364} (\bibinfo {year} {1957})}\BibitemShut
  {NoStop}%
\bibitem [{\citenamefont {Hedin}(1965)}]{Hedin1965New}%
  \BibitemOpen
  \bibfield  {author} {\bibinfo {author} {\bibfnamefont {L.}~\bibnamefont
  {Hedin}},\ }\href {https://doi.org/10.1103/PhysRev.139.A796} {\bibfield
  {journal} {\bibinfo  {journal} {Phys. Rev.}\ }\textbf {\bibinfo {volume}
  {139}},\ \bibinfo {pages} {A796} (\bibinfo {year} {1965})}\BibitemShut
  {NoStop}%
\bibitem [{\citenamefont {Mahan}(2000)}]{mahan2000many}%
  \BibitemOpen
  \bibfield  {author} {\bibinfo {author} {\bibfnamefont {G.}~\bibnamefont
  {Mahan}},\ }\href@noop {} {\emph {\bibinfo {title} {Many-body physics}}}\
  (\bibinfo  {publisher} {Kluwer Academic/Plenum Publishers, New Yoek},\
  \bibinfo {year} {2000})\BibitemShut {NoStop}%
\bibitem [{\citenamefont {Stern}(1967)}]{Stern1967}%
  \BibitemOpen
  \bibfield  {author} {\bibinfo {author} {\bibfnamefont {F.}~\bibnamefont
  {Stern}},\ }\href {https://doi.org/10.1103/PhysRevLett.18.546} {\bibfield
  {journal} {\bibinfo  {journal} {Physical Review Letters}\ }\textbf {\bibinfo
  {volume} {18}},\ \bibinfo {pages} {546} (\bibinfo {year} {1967})}\BibitemShut
  {NoStop}%
\bibitem [{\citenamefont {DuBois}(1959{\natexlab{a}})}]{DuBois1959}%
  \BibitemOpen
  \bibfield  {author} {\bibinfo {author} {\bibfnamefont {D.~F.}\ \bibnamefont
  {DuBois}},\ }\href {https://doi.org/10.1016/0003-4916(59)90016-8} {\bibfield
  {journal} {\bibinfo  {journal} {Annals of Physics}\ }\textbf {\bibinfo
  {volume} {7}},\ \bibinfo {pages} {174} (\bibinfo {year}
  {1959}{\natexlab{a}})}\BibitemShut {NoStop}%
\bibitem [{\citenamefont {DuBois}(1959{\natexlab{b}})}]{DuBois1959a}%
  \BibitemOpen
  \bibfield  {author} {\bibinfo {author} {\bibfnamefont {D.~F.}\ \bibnamefont
  {DuBois}},\ }\href {https://doi.org/10.1016/0003-4916(59)90062-4} {\bibfield
  {journal} {\bibinfo  {journal} {Annals of Physics}\ }\textbf {\bibinfo
  {volume} {8}},\ \bibinfo {pages} {24} (\bibinfo {year}
  {1959}{\natexlab{b}})}\BibitemShut {NoStop}%
\bibitem [{\citenamefont {Lee}\ \emph {et~al.}(1975)\citenamefont {Lee},
  \citenamefont {Ting},\ and\ \citenamefont {Quinn}}]{Lee1975}%
  \BibitemOpen
  \bibfield  {author} {\bibinfo {author} {\bibfnamefont {T.~K.}\ \bibnamefont
  {Lee}}, \bibinfo {author} {\bibfnamefont {C.~S.}\ \bibnamefont {Ting}},\ and\
  \bibinfo {author} {\bibfnamefont {J.~J.}\ \bibnamefont {Quinn}},\ }\href
  {https://doi.org/10.1103/physrevlett.35.1048} {\bibfield  {journal} {\bibinfo
   {journal} {Physical Review Letters}\ }\textbf {\bibinfo {volume} {35}},\
  \bibinfo {pages} {1048} (\bibinfo {year} {1975})}\BibitemShut {NoStop}%
\bibitem [{\citenamefont {Ting}\ \emph {et~al.}(1975)\citenamefont {Ting},
  \citenamefont {Lee},\ and\ \citenamefont {Quinn}}]{Ting1975}%
  \BibitemOpen
  \bibfield  {author} {\bibinfo {author} {\bibfnamefont {C.~S.}\ \bibnamefont
  {Ting}}, \bibinfo {author} {\bibfnamefont {T.~K.}\ \bibnamefont {Lee}},\ and\
  \bibinfo {author} {\bibfnamefont {J.~J.}\ \bibnamefont {Quinn}},\ }\href
  {https://doi.org/10.1103/physrevlett.34.870} {\bibfield  {journal} {\bibinfo
  {journal} {Physical Review Letters}\ }\textbf {\bibinfo {volume} {34}},\
  \bibinfo {pages} {870} (\bibinfo {year} {1975})}\BibitemShut {NoStop}%
\bibitem [{\citenamefont {Vinter}(1975)}]{Vinter1975Correlation}%
  \BibitemOpen
  \bibfield  {author} {\bibinfo {author} {\bibfnamefont {B.}~\bibnamefont
  {Vinter}},\ }\href {https://doi.org/10.1103/PhysRevLett.35.1044} {\bibfield
  {journal} {\bibinfo  {journal} {Phys. Rev. Lett.}\ }\textbf {\bibinfo
  {volume} {35}},\ \bibinfo {pages} {1044} (\bibinfo {year}
  {1975})}\BibitemShut {NoStop}%
\bibitem [{\citenamefont {Zhang}\ and\ \citenamefont
  {Das~Sarma}(2005)}]{Zhang2005}%
  \BibitemOpen
  \bibfield  {author} {\bibinfo {author} {\bibfnamefont {Y.}~\bibnamefont
  {Zhang}}\ and\ \bibinfo {author} {\bibfnamefont {S.}~\bibnamefont
  {Das~Sarma}},\ }\href {https://doi.org/10.1103/PhysRevB.72.075308} {\bibfield
   {journal} {\bibinfo  {journal} {Physical Review B}\ }\textbf {\bibinfo
  {volume} {72}},\ \bibinfo {pages} {075308} (\bibinfo {year}
  {2005})}\BibitemShut {NoStop}%
\bibitem [{\citenamefont {Hu}\ and\ \citenamefont
  {Das~Sarma}(1993)}]{Hu1993Many}%
  \BibitemOpen
  \bibfield  {author} {\bibinfo {author} {\bibfnamefont {B.~Y.-K.}\
  \bibnamefont {Hu}}\ and\ \bibinfo {author} {\bibfnamefont {S.}~\bibnamefont
  {Das~Sarma}},\ }\href {https://doi.org/10.1103/PhysRevB.48.5469} {\bibfield
  {journal} {\bibinfo  {journal} {Phys. Rev. B}\ }\textbf {\bibinfo {volume}
  {48}},\ \bibinfo {pages} {5469} (\bibinfo {year} {1993})}\BibitemShut
  {NoStop}%
\bibitem [{\citenamefont {{Das Sarma}}\ and\ \citenamefont
  {{Liao}}(2021)}]{Sarma2021know}%
  \BibitemOpen
  \bibfield  {author} {\bibinfo {author} {\bibfnamefont {S.}~\bibnamefont {{Das
  Sarma}}}\ and\ \bibinfo {author} {\bibfnamefont {Y.}~\bibnamefont {{Liao}}},\
  }\href@noop {} {\bibfield  {journal} {\bibinfo  {journal} {arXiv e-prints}\
  ,\ \bibinfo {eid} {arXiv:2101.07802}} (\bibinfo {year} {2021})},\ \Eprint
  {https://arxiv.org/abs/2101.07802} {arXiv:2101.07802 [cond-mat.str-el]}
  \BibitemShut {NoStop}%
\bibitem [{\citenamefont {Cox}\ and\ \citenamefont {Jarrell}(1996)}]{Cox1996}%
  \BibitemOpen
  \bibfield  {author} {\bibinfo {author} {\bibfnamefont {D.}~\bibnamefont
  {Cox}}\ and\ \bibinfo {author} {\bibfnamefont {M.}~\bibnamefont {Jarrell}},\
  }\href {https://doi.org/10.1088/0953-8984/8/48/012} {\bibfield  {journal}
  {\bibinfo  {journal} {Journal of Physics: Condensed Matter}\ }\textbf
  {\bibinfo {volume} {8}},\ \bibinfo {pages} {9825} (\bibinfo {year}
  {1996})}\BibitemShut {NoStop}%
\bibitem [{\citenamefont {Giuliani}\ and\ \citenamefont
  {Quinn}(1982)}]{Giuliani1982}%
  \BibitemOpen
  \bibfield  {author} {\bibinfo {author} {\bibfnamefont {G.~F.}\ \bibnamefont
  {Giuliani}}\ and\ \bibinfo {author} {\bibfnamefont {J.~J.}\ \bibnamefont
  {Quinn}},\ }\href {https://doi.org/10.1103/PhysRevB.26.4421} {\bibfield
  {journal} {\bibinfo  {journal} {Phys. Rev. B}\ }\textbf {\bibinfo {volume}
  {26}},\ \bibinfo {pages} {4421} (\bibinfo {year} {1982})}\BibitemShut
  {NoStop}%
\bibitem [{\citenamefont {Ahn}\ and\ \citenamefont
  {Das~Sarma}(2021)}]{Ahn2021}%
  \BibitemOpen
  \bibfield  {author} {\bibinfo {author} {\bibfnamefont {S.}~\bibnamefont
  {Ahn}}\ and\ \bibinfo {author} {\bibfnamefont {S.}~\bibnamefont
  {Das~Sarma}},\ }\href {https://doi.org/10.1103/PhysRevB.103.L041303}
  {\bibfield  {journal} {\bibinfo  {journal} {Phys. Rev. B}\ }\textbf {\bibinfo
  {volume} {103}},\ \bibinfo {pages} {L041303} (\bibinfo {year}
  {2021})}\BibitemShut {NoStop}%
\bibitem [{\citenamefont {Hofmann}\ \emph {et~al.}(2014)\citenamefont
  {Hofmann}, \citenamefont {Barnes},\ and\ \citenamefont
  {Das~Sarma}}]{Hofmann2014why}%
  \BibitemOpen
  \bibfield  {author} {\bibinfo {author} {\bibfnamefont {J.}~\bibnamefont
  {Hofmann}}, \bibinfo {author} {\bibfnamefont {E.}~\bibnamefont {Barnes}},\
  and\ \bibinfo {author} {\bibfnamefont {S.}~\bibnamefont {Das~Sarma}},\ }\href
  {https://doi.org/10.1103/PhysRevLett.113.105502} {\bibfield  {journal}
  {\bibinfo  {journal} {Phys. Rev. Lett.}\ }\textbf {\bibinfo {volume} {113}},\
  \bibinfo {pages} {105502} (\bibinfo {year} {2014})}\BibitemShut {NoStop}%
\end{thebibliography}%

\begin{figure*}
     \centering
     \subfloat[][Evolution of the coherent ($A_\mathrm{QP}$) and incoherent ($A_\mathrm{inc}$) parts of the spectral functions with increasing $k$ for small $r_s=0.2$]{\includegraphics[width=\linewidth]{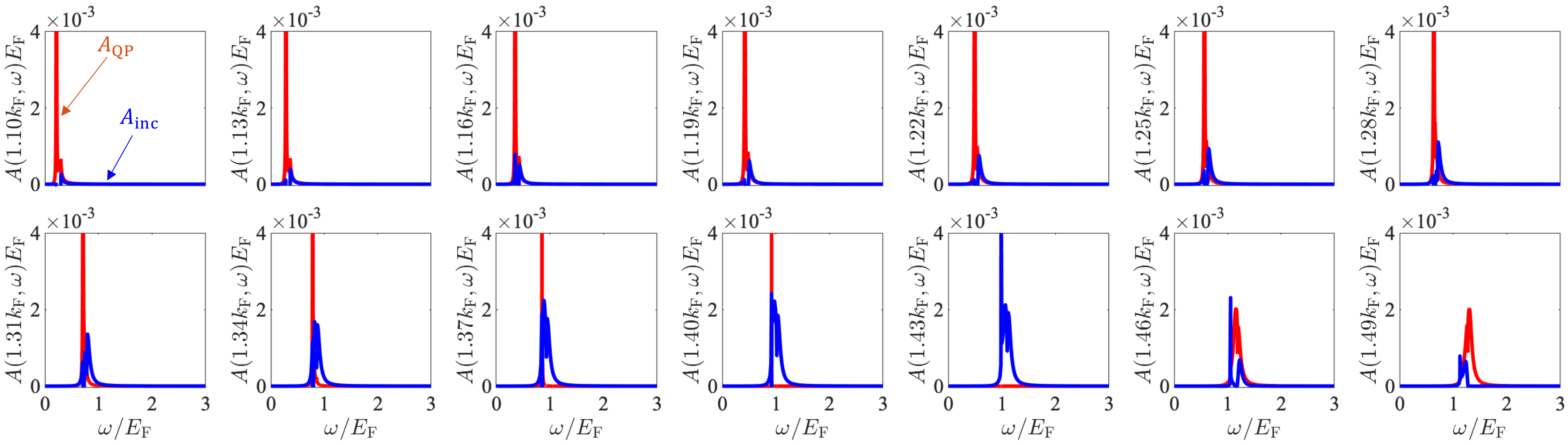}\label{fig:spectral_Aqc_Ainc_0.2}}
     
    \subfloat[][Same as (a) but for intermediate $r_s=1.0$.]{\includegraphics[width=\linewidth]{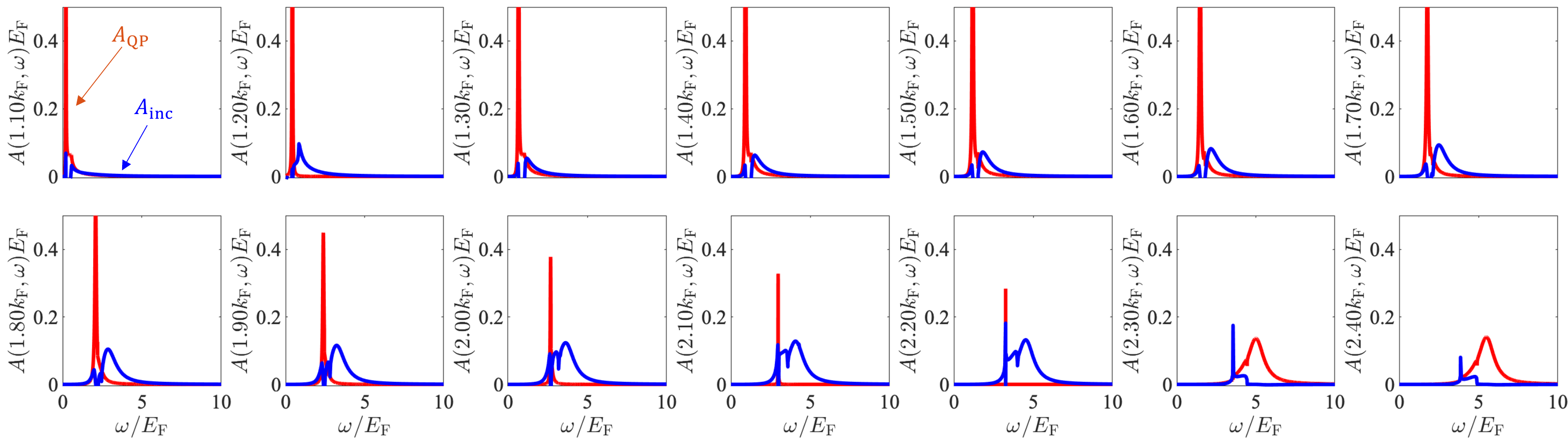}\label{fig:spectral_Aqc_Ainc_1.0}}
     
    \subfloat[][Same as (a) but for large $r_s=2.0$.]{\includegraphics[width=\linewidth]{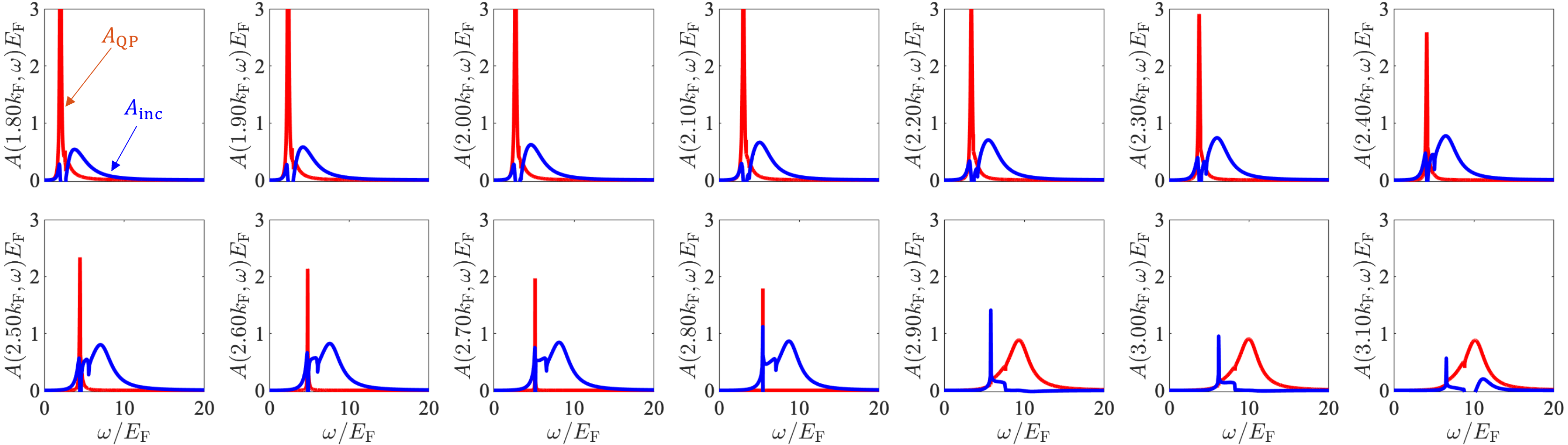}\label{fig:spectral_Aqc_Ainc_2.0}}
    \caption{Evolution of the spectral functions decomposed into the coherent and incoherent parts with increasing momentum $k$ for (a) small, (b) intermediate, and (c) large values of $r_s$.}
    \label{fig:spectral_Aqc_Ainc_Appendix}    \label{fig:spectral_Aqc_Ainc_Appendix}
\end{figure*}


\end{document}